\documentclass[12pt, usletter]{article}
\usepackage{amsmath}
\usepackage{graphicx}
\usepackage[round]{natbib}
\usepackage{url} 

\usepackage{authblk}            

\usepackage{txfonts}

\usepackage[plain,noend]{algorithm2e}

\usepackage{setspace}
\usepackage{graphicx} 
\usepackage{amsmath} 
\usepackage{amssymb}
\usepackage{subcaption}
\usepackage{float}
\usepackage{enumerate}
\usepackage{booktabs}
\usepackage{xcolor}

\usepackage[hidelinks=true]{hyperref}

\newcommand{\blind}{0}

\newcommand{\N}{\mathcal{N}}
\newcommand{\dee}{\mathrm{d}}

\newcommand{\trans}{\intercal}

\newcommand{\f}{\mathbf{f}}
\newcommand{\I}{\mathcal{I}}

\newcommand{\betabar}{\bar{\beta}}

\newcommand{\LL}{\mathcal{L}}
\newcommand{\E}{\mathrm{E}}

\definecolor{lightgrey}{HTML}{D3D3D3}
\definecolor{grey}{HTML}{BEBEBE}
\definecolor{darkorange}{HTML}{FF8C00}
\definecolor{lightslategrey}{HTML}{778899}
\definecolor{darkorchid}{HTML}{9932CC}
\definecolor{dodgerblue}{HTML}{1E90FF}
\definecolor{dodgerblue3}{HTML}{1874CD}
\definecolor{firebrickred}{HTML}{B22222}
\definecolor{vermilion}{HTML}{E34234}

\newcommand{\sw}[1]{{\color{blue} #1}}
\newcommand{\response}[1]{\textcolor{black}{#1}}

\begin{document}

\def\spacingset#1{\renewcommand{\baselinestretch}%
{#1}\small\normalsize} \spacingset{1}



\newcommand{\Title}{Model interpretation through lower-dimensional
  posterior~summarization}

\if0\blind {

  \title{\bf \Title}






  \author[1]{Spencer Woody\thanks{Corresponding author. Email to
      \texttt{spencer.woody@utexas.edu}}\footnote{ We gratefully
      acknowledge support from the Salem Center for Policy at the
      University of Texas at Austin McCombs School of Business.  }}

\author[2,1]{Carlos M. Carvalho}

\author[2,1]{Jared S. Murray
}

\affil[1]{Department of Statistics and Data Sciences, University~of~Texas~at~Austin}

\affil[2]{Department of Information, Risk, and Operations Management, University~of~Texas~at~Austin}

\maketitle

} \fi

\if1\blind
{
  \bigskip
  \bigskip
  \bigskip
  \begin{center}
    {\LARGE\bf \Title} \\ \medskip
    {\today}
\end{center}
  \medskip
} \fi


\bigskip
\begin{abstract}                
  Nonparametric regression models have recently surged in their power
  and popularity, accompanying the trend of increasing dataset size
  and complexity.  While these models have proven their predictive
  ability in empirical settings, they are often difficult to interpret
  and do not address the underlying inferential goals of the analyst
  or decision maker.  In this paper, we propose a modular two-stage
  approach for creating parsimonious, interpretable summaries of
  complex models which allow freedom in the choice of modeling
  technique and the inferential target.  In the first stage a flexible
  model is fit which is believed to be as accurate as possible.  In
  the second stage, lower-dimensional summaries are constructed by
  projecting draws from the distribution onto simpler structures.
  These summaries naturally come with valid Bayesian uncertainty
  estimates.  Further, since we use the data only once to move from
  prior to posterior, these uncertainty estimates remain valid across
  multiple summaries and after iteratively refining a summary.  We
  apply our method and demonstrate its strengths across a range of
  simulated and real datasets.  Code to reproduce the examples shown
  is avaiable at \texttt{github.com/spencerwoody/ghost}.
\end{abstract}

\noindent%
 \vfill


\newpage
\spacingset{1.5} 


\section{Introduction}
\label{sec:introduction}

In regression modeling, there has traditionally existed a natural
tension between model flexibility and interpretability.  Consider the
generic regression model given by
\begin{align}\label{eq:generic-regression}
  \E[y \mid x] = f(x). 
\end{align}
There are many models available to estimate the function $f$, which
describes the relationship between the covariates $x$ and the expected
outcome of the noisy observations $y$.
On one hand, simple models such as the linear model or a shallow
regression tree are readily interpretable, but are likely biased
because they cannot capture complex relationships between the input
and the response.  On the other hand, more complex nonparametric
regression models can yield highly accurate predictions but are
difficult to interpret.  

In particular, we often would like to answer questions such as: which
covariates have the strongest effect on prediction?  Does covariate
importance differ across the covariate space?  Are there interactions
among the covariates, and if so, which are most important?  Answering
such questions is difficult, and providing appropriate measures of
uncertainty is even more so.

In this paper, we propose an approach to give interpretable model
summaries designed to answer such questions.  We assume a Bayesian
vantage point throughout, so that a flexible prior is specified for
the regression function $f$ and the posterior is calculated by
conditioning on observed data. The key idea of our approach is to
follow a two-stage process.  First, specify a flexible model for $f$
and use all the available data to best estimate this relationship.
Second, perform a \textit{post hoc} investigation of the fitted model
using lower-dimensional surrogates which are suited to answer relevant
inferential questions and sufficiently representative of the model's
predictions.  These summaries are functions of $f$, so obtaining their
posterior distribution is straightforward.  This investigation in the
second stage is simply an exploration of the posterior for $f$.  The
result is a set of interpretable explanations of model behavior, along
with posterior distributions for these explanations, which are valid
in the sense that we condition on the data only once (in calculating
the posterior for $f$).

\response{The methods we develop are highly modular, allowing for a
  variety summaries to interpret many possible regression models.
  Here we demonstrate three particular uses of posterior summarization
  which we believe to be widely advantageous and represent valuable
  contributions, including (i) to efficiently and intuitively describe
  partial effects, (ii) to communicate the most significant
  interactions detected by the first-stage regression model, and (iii)
  to describe how the predictive importance of covariates changes
  across local areas of the covariate space.  Because we take a
  Bayesian approach, these summaries all come with posterior
  uncertainty estimates.  }



The remainder of the paper is structured as follows.  In the rest of
this section we develop our decision theoretic approach for producing
model summaries, and describe several metrics for gauging the how well
model summaries explain model predictions.  In
Section~\ref{sec:high-dimens-line} we consider the case of estimating
and quantifying posterior uncertainty in lower-dimensional summaries
for linear models.  This leads us into Section~\ref{sec:nonp-regr}
where we generalize this approach to summarize nonparametric
regression models.  In Section~\ref{sec:simul-results-nonp} we present
simulation results which show how our method of nonparametric
regression summaries can accurately communicate the average partial
effects of covariates and convey significant interactions present in a
fitted regression model.  We apply our method to a real data example
in Section~\ref{sec:exampl-calif-hous} by presenting an extensive case
study interpreting a predictive model for housing prices in
California.  We conclude with a discussion in
Section~\ref{sec:discussion}.



\subsection{Separating modeling and interpretation}
\label{sec:separ-model-infer}

Our goal is to produce interpretable summaries of regression models.
Equivalently, we wish to understand important predictive trends in the
regression function $f$ from Eq.~\eqref{eq:generic-regression}.  To do
so, we consider a lower-dimensional class of functions $\Gamma$ that
can be used for parsimoniously characterizing $f$.  We can summarize
$f$ by finding a $\gamma \in \Gamma$ that closely matches its
predictions.  Formally, we let $\gamma$ be the function minimizing an
objective defined by
\begin{align}\label{eq:general-loss-fn}
  \LL(f, \gamma, \tilde{X}) &= d(f, \gamma, \tilde{X}) + q_\lambda(\gamma),
\end{align}
where $d(\cdot, \cdot, \tilde{X})$ measures discrepancy in prediction
between the original high-dimensional model $f$ and the parsimonious
summary $\gamma$ over some $\tilde n$ specified covariate locations of
interest $\tilde X$, and $q_\lambda(\cdot)$ is an optional penalty
function measuring complexity in $\gamma$ governed by one or several
tuning parameters $\lambda$.  The penalty $q_\lambda(\cdot)$ may be
used, for instance, to enforce sparsity or smoothness in the summary.
Now the summary is the function minimizing this objective,
\begin{align}
  \label{eq:functional}
  \gamma(x) = \arg \min_{\gamma' \in \Gamma} \LL(f, \gamma', \tilde X).
\end{align}
Of course, we do not know the true regression function $f$, but rather
have a posterior distribution for it.  Because $\gamma$ here is a
functional of the regression function $f$, it has a posterior implied
by that of $f$.  That is, the posterior of $\gamma$ is precisely the
posterior distribution of the best approximation of $f$ in the class
$\Gamma$ as measured by the penalized predictive discrepancy in
\eqref{eq:general-loss-fn}.

For example, if $\Gamma$ is the set of linear functions and there is
no penalty $q_\lambda(\cdot)$, then we obtain the posterior
distribution of the best linear approximation to $f$, thereby
describing the average partial effects of covariates on the
conditional expectation of the outcome.  We obtain this directly,
without fitting a misspecified linear model for the outcome $y$ from
the outset.  We can also simultaneously consider linear summaries in
$k < p$ variables, additive summaries, and so on, all with valid
Bayesian inference.

The summary objective \eqref{eq:general-loss-fn} is flexible by
design, allowing $\tilde{X}$ to be any set of chosen covariate
locations, possibly with different weights assigned within the
discrepancy function $d(\cdot, \cdot, \tilde X)$.  If it is chosen to
be the entire dataset, then the result is a global summary of model
predictions.  If $\tilde X$ is a subset of the data confined to a
restricted region, the result is a local summary of model predictions
within this region.  This is particularly helpful, as nonparametric
regression models naturally adapt to local heterogeneity in covariate
importance; for instance, in Section~\ref{sec:local-linear-summary} we
illustrate how the determinants of housing prices vary geographically.
If $\tilde X$ is chosen to be a set of locations where the outcome has
not been observed, then the summary explains how the model makes
predictions at these new locations.

This distribution for $\gamma$ given by \eqref{eq:functional} accounts
for uncertainty in the summary function, but leaves open the question
of a point estimate.  Using standard Bayesian decision theory
\citep[e.g., ][]{berger2013statistical}, if we cast the objective
function \eqref{eq:general-loss-fn} as a loss function then the
optimal point estimate for the summary is that which minimizes
posterior expected loss, i.e.
\begin{align}\label{eq:general-loss-min}
  \hat{\gamma}(x) &:= \arg \min_{{\gamma'} \in \Gamma}
                    \E[\LL(f, \gamma', \tilde{X}) \mid Y, X],
\end{align}
with this expectation taken over $f$.  When
$d(\cdot, \cdot, \tilde{X})$ is chosen to be squared-error, then the
point summary is equivalent to minimizing the loss function in
\eqref{eq:general-loss-fn} with the posterior mean $\hat{f}$ taking
place of $f$, and so the point estimate becomes
\begin{align}
  \hat{\gamma}(x) &= \arg \min_{\gamma' \in \Gamma}
                    \tilde{n}^{-1} \sum_{i=1}^{\tilde n} [\hat f(\tilde
                    x_i) - \gamma'(\tilde x_i)]^2 
                    +
                    q_\lambda(\gamma). \nonumber
\end{align}
This form conveniently lends itself to the use of many standard
estimation procedures, where the summary estimate $\hat \gamma$ is
obtained from the fitted values of $\hat f$.  The tuning parameter(s)
$\lambda$ may be selected using usual approaches adapted for this
case, e.g. using cross-validation on the values of
$\hat f(\tilde x_i)$.

\response{Here we pause to emphasize the subtle distinction between
  the point estimate $\hat \gamma$ in \eqref{eq:general-loss-min} and
  the posterior for the summary given by \eqref{eq:functional}.  The
  former is the point estimate (more precisely, the Bayes estimate) of
  the summary, while the latter is the entire posterior for the
  lower-dimensional characterization of $f$.  This is akin to the
  difference between the Bayes estimator for a scalar parameter, and
  the posterior for that parameter. The posterior mean of
  $p(\gamma \mid Y)$ is not necessarily equivalent to $\hat{\gamma}$,
  i.e. in general
  $\arg \min_\gamma \E[\LL (f, \gamma, \tilde X) \mid X, Y]
    \neq
    \E [\arg \min_\gamma \LL (f, \gamma, \tilde X)\mid X, Y].$
  }

  Our work is related to that of
  \cite{crawford2018kernel,crawford2019predictor} who calculate linear
  projections of nonlinear regression models to produce an ``effect
  size analog'' for each covariate.  Here, however, we allow for a
  more general set of possible summaries, and introduce heuristics to
  iterative update a summary to produce a more faithful representation
  of the regression model.  Additionally, we extend and generalize
  previous approaches which derive decision-theoretic point estimate
  model summaries; for example, \cite{DSS} introduce posterior
  summarization for communicating dominant trends in linear models.
  This framework has been shown to be effective in a variety of
  modeling contexts \citep{PuelzSUR, Chakraborty2016,optimalETF,
    PuelzPortfolio, bashir2018post, kowal2018bayesian,
    maceachern2019economic, lee2014inference}.  Related ideas in this
  direction can be traced back to \cite{maceachern2001decision} who
  developed linear summaries for nonparametric regression models.

\subsection{Summary diagnostics}
\label{sec:summary-diagnostics}

A natural concern after summarization is the adequacy of the summary
function approximation to the regression function.  The summary will
generally have less predictive power than $f$ because it sacrifices
flexible predictive features in $f$ such as nonlinearities or
interactions.  There are several ways one may gauge this.

We propose two diagnostic metrics to quantify the sufficiency of
summarization.  The first measures predictive variance in the original
model explained by summarization, 
\begin{align}
  R^2_\gamma
  &:= 1 - \frac
    {\sum_i [f(\tilde x_i) - \gamma(\tilde  x_i)]^2}
    {\sum_i [f(\tilde x_i) - \bar{f}]^2},
    \nonumber
\end{align}
where $\bar{f} := \tilde n^{-1}\sum_i f(\tilde x_i)$.  This is the
``summary $R^2$.'' The second metric, which can be used for the case
of 
normal errors, is
\begin{align}
  \phi_\gamma &= {\sqrt{ \tilde n^{-1}
                \sum_i 
                \left[
                \tilde y_i - \gamma( \tilde x_i)
                \right]^2} / \sigma } -1
                \nonumber
\end{align}
where $\tilde y_i$ is the observation corresponding to $\tilde x_i$.
This metric has the loose interpretation that using the summary model
increases the width of predictive intervals by
$(\phi_\gamma \times 100)\%$.  If the observations $\tilde y_i$ are
not available, then we can use estimates from the posterior predictive
$p(\tilde y_i \mid Y, X, \tilde x_i)$.  Similar quantities may be
calculated for non-normal errors.  Both of these metrics also come
with posterior distributions, calculated by using posterior draws of
$f$, $\gamma$, and $\sigma$.

Furthermore, one can visually inspect the ``summary residuals''
$\hat f(\tilde x) - \hat \gamma(\tilde x)$, either with a scatter plot
or fitting a single regression tree, which could reveal important
interactive effects in $f$ driving variation in the summary residuals
that should be considered.

After analyzing the summary model in this way, either quantitatively
with these two metrics or qualitatively through the summary residuals,
we may be determine that the class of summaries was too simplistic to
satisfactorily explain the original model. Then it is appropriate to
specify a more nuanced class of summary, such as one which allows for
interactions, or one that allows for nonlinear rather than linear
effects.  This suggests an iterative approach of progressively
assessing and updating the class of summaries until one or more
summaries is deemed to be sufficiently representative of the original
model's predictions.  Critically, our summarization and posterior
projection approach still yields valid Bayesian inference after this
``summary search.''  \response{We detail this iterative approach in
  Section~\ref{sec:nonp-regr}. }



\section{Lower-dimensional summaries in the linear model}
\label{sec:high-dimens-line}

We first consider the relatively simple case of summarizing a
high-dimensional linear model with a subset of the variables.  We
extend the work of \cite{DSS} by introducing measures of uncertainty
in the summary via posterior projection.

The full model is a standard multiple linear regression,
$(y \mid \beta, \sigma^2) \sim \N(X\beta, \sigma^2\I)$, with
independent priors
$\pi(\beta, \sigma^2) = \pi(\beta) \cdot \pi(\sigma^2).$
We wish to find a sparse set of relevant features.  Denote this set by
the inclusion vector $\eta \in \{0, 1\}^p$.  Using the notation
introduced in the previous section, this is equivalent to replacing
the original fitted function $f(x) = x^\trans \beta$ with the summary
$\gamma(x) = x^\trans \tilde\beta$ for a sparse vector $\tilde\beta$,
where $\tilde \beta_j = 0$ if $\eta_j = 0$.  If we use the
squared-error function to measure predictive discrepancy and some
sparsity-enforcing penalty $q(\tilde\beta)$ \citep[such as the
$\ell_1$ penalty of][]{lasso} then the optimal sparse summary point
estimate is
\begin{align}\label{eq:dss-point}
  \beta_\lambda := \arg \min_{\tilde\beta}
  N^{-1} \| X\betabar - X\tilde\beta \|_2^2 + 
  \lambda \cdot q(\tilde\beta)
\end{align}
where $\bar\beta$ is the posterior mean of $\beta$.  Note that this
matches Eq. (20) of \cite{DSS}.  The penalty term
$q(\tilde\beta)$ is included solely for sparsity in the solution.  For
any such penalty, \eqref{eq:dss-point} returns an entire solution path
for possible sparse summaries of the original high-dimensional model,
with the level of sparsity varying with the tuning parameter
$\lambda$.

After solving \eqref{eq:dss-point} for some fixed value of $\lambda$,
we have sparse set of coefficients which is a Bayes-optimal point
estimate summary for the full model.  Using our posterior projection
technique, we can also quantify uncertainty in this summary.  A naive
approach would be to refit the model only with the selected
covariates.  However, this would involve using the outcome data $y$ a
second time---an example of ``posterior hacking,'' or
opportunistically retraining a new model after already conditioning on
the data once in the original model.

Instead, it is more appropriate to propagate posterior uncertainty
from the original fitted model through to the linear summary.  The
sensible way to do this is to take the full posterior distribution for
the fitted function of the full model using all the variables, and
project it onto the space of the fitted summary function using the
restricted set of variables.  We use the data exactly once (in
obtaining the posterior for the original full model) and obtain the
posterior of the best linear approximation in $k < p$ variables.

To be more specific, for one value of $\lambda$, denote the
corresponding sparse model summary with the inclusion vector
$\eta_\lambda$, whose $j$\textsuperscript{th} element is 0 if
$(\beta_\lambda)_j = 0$ and 1 otherwise.
Given a sparse linear summary specified by $\eta$ (for notational
simplicity, dropping the $\lambda$ subscript), we want to give a
coherent posterior distribution to the included coefficients.  This is
the posterior for the low-dimensional linear representation of the
original model.

Let $X_\eta$ denote the $\eta$-subset of the columns of the original
covariate matrix $X$, and let $\beta_\eta$ be the vector of
coefficients for this restricted covariate matrix.  We wish to map the
posterior for $X\beta$, the original fitted values, onto
$X_\eta \beta_\eta$, the fitted values using the restricted set of
coefficients.  This is equivalent to projecting the original fitted
values $X\beta$ onto the column space of $X_\eta$.  We can approximate
the posterior distribution $p(\beta_\eta \mid y)$ for the restricted
covariates via Monte Carlo, i.e., for the $k$th draw from the original
posterior, $\beta^{(k)} \sim p(\beta \mid y)$, perform the projection
\begin{align}\label{eq:linear-projection}
  \beta_\eta^{(k)} = (X^\trans_\eta X_\eta)^{-1}X_\eta^\trans X\beta^{(k)},
\end{align}
assuming the inverse exists.  For this reason, we call the
$p(\beta_\eta \mid y)$ the ``projected posterior.''

In this way we can obtain projected posteriors for all sparse summary
models from the solution path given by \eqref{eq:dss-point}, and
report the summary which is sufficiently representative of the
original full model's predictions, as measured by the summary
diagnostic measures given in Section~\ref{sec:summary-diagnostics}.
We emphasize that $\beta_\lambda$ in \eqref{eq:dss-point} is the
Bayes-optimal point estimate for the summary, and the projected
posterior represents posterior uncertainty around this estimate.


\subsection{Sparse linear summaries for the US crime data}
\label{sec:example:-us-crime}

Here we illustrate our approach of quantifying uncertainty in sparse
linear summaries on the US crime dataset, which has $n=47$
observations with $p=15$ predictors.  We fit a linear model using the
horseshoe prior \citep*{horseshoe} after log-transforming the
continuous variables, and centering and scaling all variables.  Then
we obtain point estimates for linear summaries of the full model by
solving the minimization problem in \eqref{eq:dss-point}.  Because the
posterior mean $\bar\beta$ is already a shinkage estimator due to the
influence of the prior, we use the adaptive lasso penalty
\citep{adaptivelasso} for the penalty term,
$q(\tilde\beta) = \sum_j w_j^{-1} |\tilde \beta_j|$ with
$w_j =|\bar\beta_j|$ to alleviate the problem of ``double shrinkage''
that would result from using the usual (unweighted) $\ell_1$ penalty.
These summaries were calculated using the lasso implementation from
the \texttt{lars} package \citep{lars} in the \textsf{R} programming
language \citep{R}.  For each point estimate summary, we calculate its
projected posterior following \eqref{eq:linear-projection}.

\begin{figure}[!htb]
  \centering
  \includegraphics[width=0.65\textwidth]{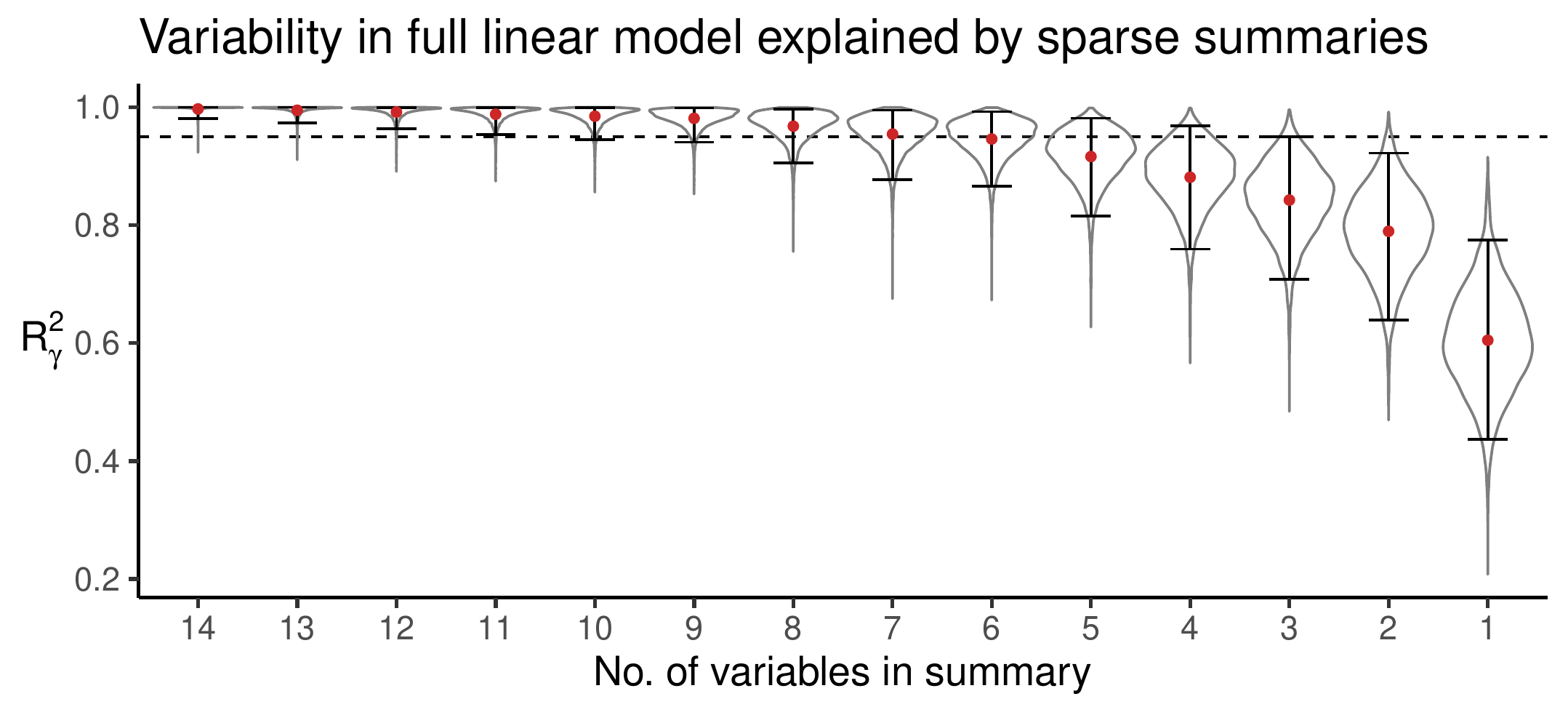}\\
  \includegraphics[width=0.65\textwidth]{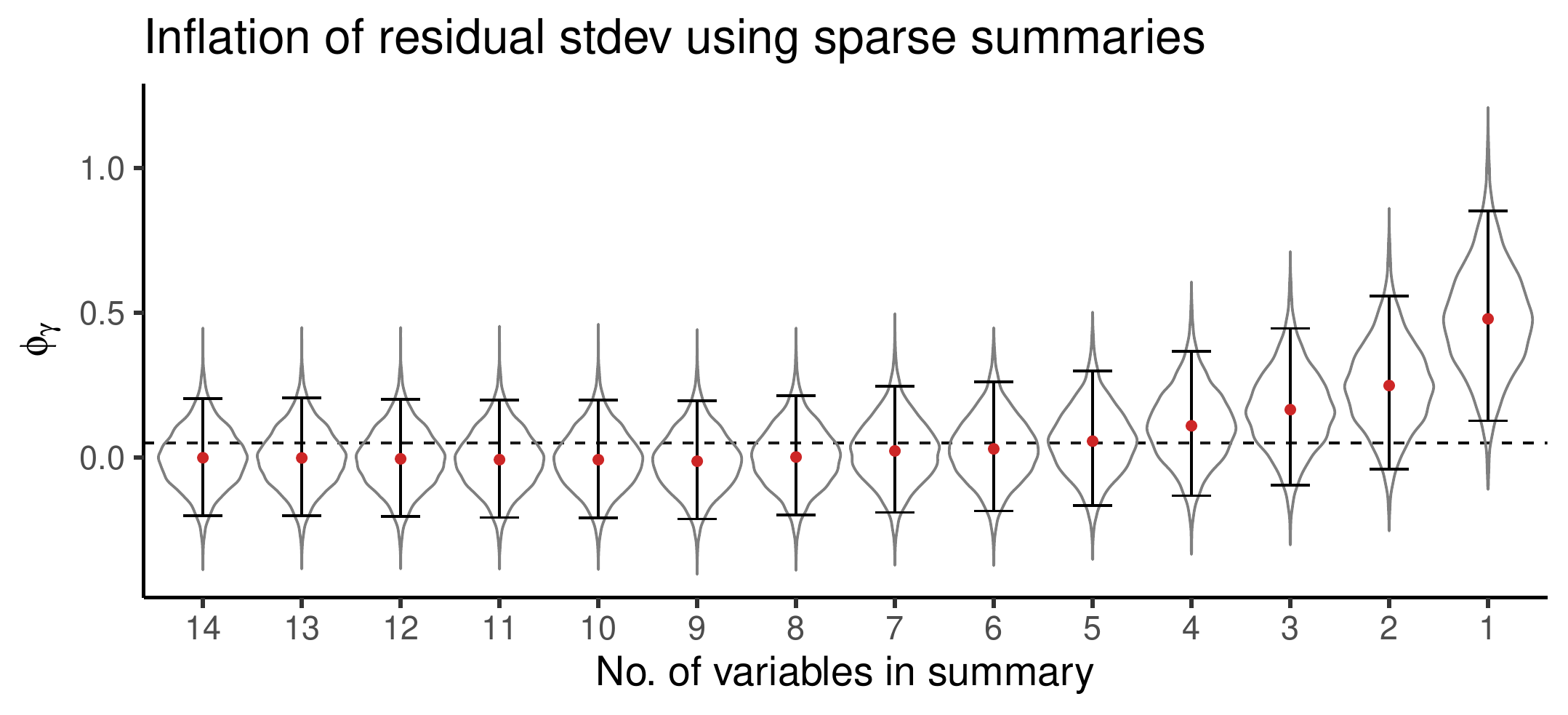}
  \caption{\label{fig:dss-crime-summaries} Diagnostics for
    low-dimensional linear summaries of crime data using horseshoe
    prior.  We recommend to report the summary model with 6
    predictors, as this summary explains about $R^2_\gamma = 95\%$ of
    predictive variation, while predictive intervals are inflated by
    only around $\phi_\gamma = 5\%$, with these values represented by dashed lines.  
  }
\end{figure}

Figure~\ref{fig:dss-crime-summaries} shows posterior distributions for
the summary diagnostics defined in
Section~\ref{sec:summary-diagnostics} for the entire solution path of
sparse linear summaries.  Following \cite{DSS}, we recommend reporting
the summary model with 6 predictors included, as this summary explains
approximately 95\% of predictive variation in the full model, and
predictive intervals are inflated by only about 5\% on average.
However, the summary diagnostics allow an analyst to pick any
reasonable tradeoff between parsimony and predictive ability, and we
can get valid inference for any summaries of interest.

\begin{figure}[!htb]
  \centering
  \includegraphics[width=0.65\textwidth]{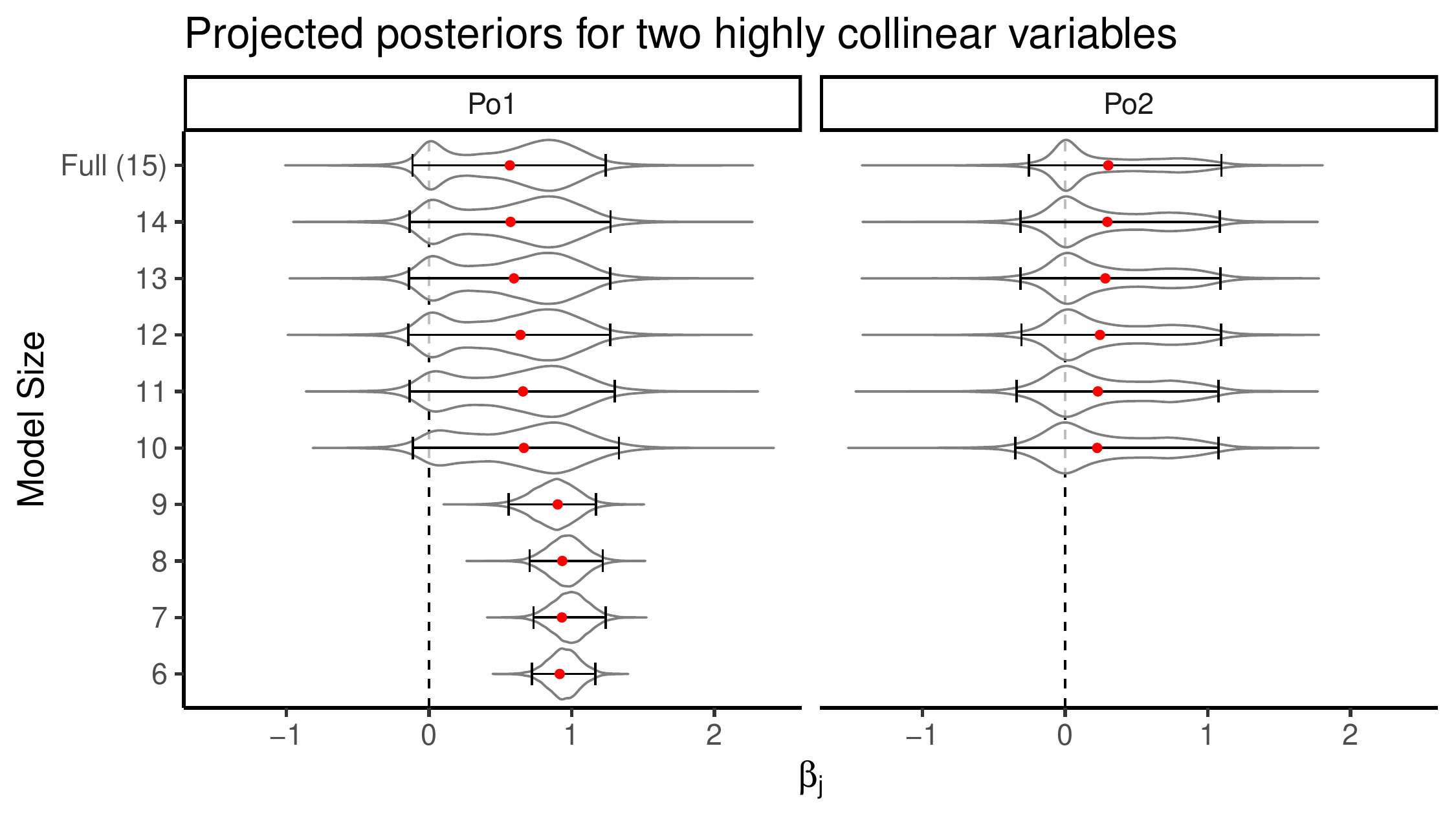}
  \caption{\label{fig:dss-crime-posteriors} Point estimate summaries
    and projected posteriors Po1 and Po2, two highly correlated
    variables in the US crime dataset.  Once Po2 is ``selected out''
    of the summary model, all the predictive power jointly
    attributable to Po1 and Po2 is shifted to Po1.  }
\end{figure}

We use this example to consider the effect of sparsification on the
shape of projected posteriors.  Figure~\ref{fig:dss-crime-posteriors}
investigates the projected posteriors for two highly collinear
variables, Po1 and Po2, as the linear summary becomes more
parsimonious.  The presence of collinearity results in both covariates
having high posterior variance in the full model, and due to the
nature of the horseshoe prior which aggressively shrinks variables
near zero while retaining heavy tails, both marginal posteriors are
bimodal with modes near and away from zero.  However, moving from the
summary with 10 variables to the summary with 9 variables (when Po2 is
``selected out'' of the summary), the projected posterior mode for Po1
near zero disappears, and all the mass in the posterior is shifted to
the right.  This shows the gain in power from using our summarization
approach.  Projected posteriors for all variables for all summaries
shown in the supplement. 

\begin{figure}[ht]
  \centering
  \includegraphics[width=0.67\textwidth]{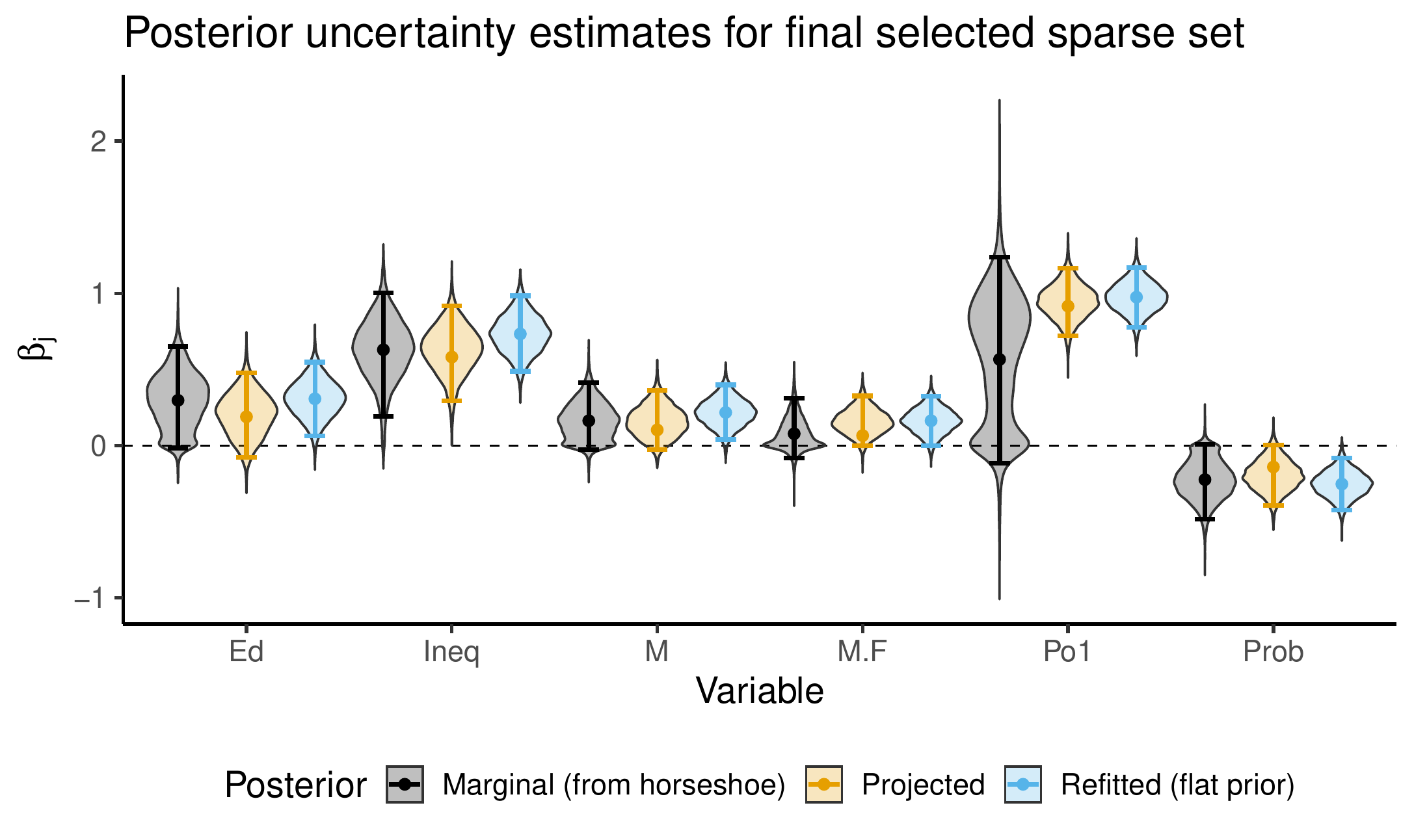}
  \caption{\label{fig:crime-uncertainty-intervals} Comparing projected
    posteriors for variables in the selected model to the posterior
    from refitting the model with the selected variables (``posterior
    hacking'') using a flat prior, and the marginal posteriors from
    the original fitted model.  The projected posteriors have wider
    credible intervals compared to the refitted posteriors,
    demonstrating propagation of model uncertainty in the original
    fitted model, and also shrink posterior means closer to zero
    compared to the refitted model.  }
\end{figure}

Finally, in Figure~\ref{fig:crime-uncertainty-intervals} we compare
the projected posterior for the final selected sparse summary model to
the posterior we would obtain by refitting the linear model only
including these variables, instead of projecting the posterior
draws. For this case we now use a flat prior on (the restricted
vector) $\beta$ as we suspect that there is less need for shrinkage
since we have reduced the dimensionality.  In this second case, we are
``double dipping'' with the data, using it once to fit the full model,
and then using it a second time after the sparse linear summary is
chosen.  This inference is not strictly valid, since the data are used
here to set the prior by selecting the restricted set of variables.
More importantly, this posterior entirely ignores model uncertainty.
By comparison, the projected posterior uses the response variable $y$
only once, in calculating the posterior for the full model.  We also
show the marginal posteriors from the original fitted (saturated)
linear model.

The projected posteriors are wider than the refitted posteriors, due
to the propagation of model uncertainty.  The projected posteriors
retain the shrinkage properties of the original posterior: the
posterior means for each variable are shrunk toward zero compared to the
refitted posterior.  In most cases the projected posteriors
closely match the marginal posteriors of the full model, the biggest
exception being Po1 for the reasons just discussed.


\section{Summaries for nonparametric regression}
\label{sec:nonp-regr}

We now move to our main topic: summarization of nonparametric
regression models.  This problem is more nuanced than for the linear
model, where increasing summary complexity was well defined by the
dimension of the sparse linear summary.  Here, however, it is less
clear how to define a collection of increasingly complex summaries
from which to choose.  This suggests an iterative approach, whereby an
initial summary is proposed, calculated, evaluated and updated as
necessary.

Before an presenting investigative simulation examples in
Section~\ref{sec:simul-results-nonp} and an extensive case study in
Section~\ref{sec:exampl-calif-hous}, we describe heuristics for model
summary search.  This is given in detail below, with
Algorithm\ref{al:1} containing a brief outline of our procedure.  The
exposition is intentionally general, meant to allow for any class of
regression models $f$ with any error distribution, and any class of
lower-dimensional summaries corresponding to inferential goals of
interest.  We also describe how this approach can be used to elucidate
how the model predicts globally or locally.  Exact details of how to
processed will be heavily context dependent, influenced by the
specific dataset, original specified model, and the inferential
target.


\subsection{Iterative summary search}
\label{sec:generic-description}

\paragraph{(1) Specify and fit the full model.  }

Assume the regression setting described by
$$\E[y \mid x] = f(x)$$ and complete the model specification by
assigning priors for the regression function $p(f)$ as well as any
nuisance parameters.  Our approach is agnostic to the choice of
$p(f)$, though we do assume that it fits well by adequately modeling
the response $y$ as a function of the covariates $x$.  Typically this
should be a nonparameteric prior, such as a Bayesian tree ensemble
\citep{chipman2010} or some variant of a Gaussian process
\citep[e.g.][]{GramacyLee2008, GramacyApley}.

We obtain $M$ Monte Carlo draws targeting the posterior of $f$,
denoted by $\{f^{(k)}\}_{k=1}^M$.  Denote the posterior mean for the
fitted value of the function at $x_i$ by
$\hat{f}(x) := M^{-1}\sum_k f^{(k)}(x) $.

\paragraph{(2) Summarize.  }

Choose a class of summaries $\Gamma$ which matches the inferential
goal at hand.  For example, if the objective is to comment on the
partial linear effect of each covariate, then $\Gamma$ is chosen to be
linear.  If instead the goal is to simply comment on the partial
effect of each covariate, without the constraint of linearity, then
one can choose $\Gamma$ to be the broader class of additive functions.

We also need to specify the predictive locations $\tilde{X}$ at which
to summarize the model output, a metric $d(\cdot, \cdot, \tilde{X})$
for measuring predictive discrepancy between the summary and the full
model, and an appropriate summary complexity penalty function
$q_\lambda(\gamma)$.  These components collectively define the
summarization loss function
\begin{align*}
  \LL(f, \gamma, \tilde{X}) &= d(f, \gamma, \tilde{X}) + q_\lambda(\gamma).
\end{align*}

The optimal point summary is
$$\hat\gamma(x) = \arg \min_{{\gamma} \in \Gamma} \E[\LL(f, \gamma,
\tilde{X}) \mid Y, X],$$ found by minimizing the summarization loss in
expectation over the posterior for $f$.  Tuning parameters can be
determined, for example, through cross-validation or use of
information criteria on the posterior mean fitted values
$\hat f(\tilde x_i)$.  Once $\hat{\gamma}(x)$ has been calculated, the
posterior distribution for the summary can be found by the posterior
of the functional
$\gamma(x)=\arg \min_{\gamma \in \Gamma} \LL(f, \gamma, \tilde X)$.
Often this will involve projecting posterior draws of the fitted
values $f^{(k)}(\tilde x_i)$ onto the predictive space of
$\hat \gamma$.



\paragraph{(3) Evaluate. } 

Next, assess the impact of moving from the full model to the
low-dimensional summary.  The summarization metrics $R^2_\gamma$ and
$\phi_\gamma$ defined in Section~\ref{sec:summary-diagnostics} offer
two readily interpretable ways to quantify this loss in predictive
power.  One may also inspect the summarization residuals,
$\hat{f}(\tilde x_i) - \hat \gamma(\tilde x_i)$ directly, for example
by training a regression tree to these residuals to detect and
characterize heterogeneity.

\paragraph{(4) If the summary is sufficient, perform inference.}

Based on the results from Step (3), determine whether the summary
model is sufficient.  For example, if $R^2_\gamma$ is reasonably high
and the summarization residual regression tree does not detect large
amounts of residual heterogeneity, then the calculated summary in Step
(2) may be judged to be of good quality and representative of the
model's predictions, and so this summary may be used for the inference
stage.  Ultimately it is left to the end user to make a decision
regarding sufficiency of the calculated model summary.

\paragraph{(5) Otherwise, refine and return to (2). } 

If the summary is deemed to be of poor quality, there are two ways to
improve model summary accuracy: the class of summary models $\Gamma$
can be enriched to allow for greater flexibility, or the predictive
locations $\tilde{X}$ can be altered to be more localized.  The choice
of one or both of these options can be informed by the evidence
provided from the evaluation procedure in Step (3).  For instance, if
the regression tree detects high levels of heterogeneity in the
summarization residuals, one may allow for low-order interactions
determined by splitting rules near the root of the tree.

With these new classes of summaries, and/or designated predictive
locations, return to Step (2) to calculate the summary and iterate
through all these steps until a summary is deemed sufficient or it is
judged that no summarization class can be specified that is
representative enough of the model's predictions while still being
interpretable.  We need not constrain ourselves to a single model
summary, however; we may compute multiple summaries to interpret of
model behavior, and these will all have valid Bayesian posteriors.

\begin{algorithm}[ht!]
  \footnotesize
  \SetKwInOut{Input}{Input}
  \SetKwInOut{Output}{Output}
  
  \Input{Outcome vector $Y$; covariate matrix $X$. }

  \begin{enumerate}[(1)]
    \item Specify and fit the model
    \begin{itemize}
    \item Give a prior $p(f)$ for $f(x) = \E[y \mid x]$.
    \item Obtain posterior samples for the model,
      $f^{(k)} \sim p(f \mid Y)$
    \item Choose an initial class of summaries $\Gamma$, complexity
      penalty $q_\lambda(\cdot)$ (if desired) and predictive locations
      $\tilde X$, collectively defining the summary loss 
      $\LL(f, \gamma, \tilde{X})$.
    \end{itemize}
  \item Summarize 
    \begin{itemize}
    \item Calculate the optimal point estimate for the
      summary
      $$\hat\gamma(x) = \arg \min_{{\gamma} \in \Gamma} \E_{f}[\LL(f,
      \gamma, \tilde{X}) \mid Y, X].$$
\item Calculate the projected posterior for the summary using
  posterior draws of $f^{(k)}\sim p(f\mid Y)$ 
  $$\gamma^{(k)}(x)=\arg \min_{\gamma \in \Gamma}\LL(f^{(k)}, \gamma, \tilde{X}).$$
\end{itemize}
  \item Evaluate
    \begin{itemize}
    \item Compute summarization metrics $R^2_\gamma$ and $\phi_\gamma$
    \item Inspect the summarization residuals with a decision tree
    \end{itemize}
  \item Iterate if necessary
    \begin{itemize}
    \item If the summary $\gamma$ is deemed satisfactory by metrics in
      (3), use this summary for interpretation.
    \item Otherwise, change the class of summaries $\Gamma$ and/or
      predictive locations $\tilde X$, and iterate through (2) and (3)
    \end{itemize}

  \end{enumerate}

\BlankLine

  \Output{Summary point estimate $\hat \gamma$ and its projected
    posterior $p(\gamma \mid Y)$.} \BlankLine
  \normalsize
  \caption{Outline of iterative procedure for summarizing
    nonparametric regression models in
    Section~\ref{sec:nonp-regr}.}\label{al:1}
\end{algorithm}\DecMargin{1em}


\section{Simulation results}
\label{sec:simul-results-nonp}

\subsection{Estimating and visualizing partial effects}
\label{sec:toy-exampl-nonadd}

Here we present a toy example to illustrate how our approach can be
used to estimate partial effects as a summarization of a nonparametric
regression model.  We simulate data from the model
\begin{align*}
  y = f(x_{1}, x_{2}) + \varepsilon, \quad \varepsilon \sim \N(0, \sigma^2)
\end{align*}
centered on the bivariate nonadditive function defined by
\begin{align*}
    f(x_1, x_2) &= {1}/\{1 + \exp(-2x_1 - 2x_2)\} + {1}/\{1 + \exp(-x_1 + 4x_2)\},
\end{align*}
with $\sigma^2 = 0.25$, and $n = 2500$ observations along a
$50 \times 50$ regular 2D grid of $(x_1, x_2)$ values over the range
$(-2, 2)$.  \response{No other covariates are observed or used to
  generate the data.  }  Using these data we estimate $f(x_1, x_2)$ by
using a Gaussian process (GP) prior with a squared exponential
covariance kernel, and assign Jeffreys' prior for $\sigma^2$.

We consider two summaries to explain model predictions in the GP
posterior for $f(x_1, x_2)$ by estimating the partial effect of each
covariate.  The first is a linear summary, so the class of summaries
$\Gamma_1$ is the set of functions of the form
$\gamma_1(x_1, x_2) = \alpha_1 + \beta_1 x_1 + \beta_2 x_2$.  The
second is an additive summary, so the class of summaries $\Gamma_2$ is
the set of functions of the form
$\gamma_2(x_1, x_2) = \alpha_2 + h_1(x_1) + h_2(x_2)$, with $h_1$ and
$h_2$ being univariate functions whose forms we discuss in the
proceeding paragraph.  Here and throughout the paper, we use the
squared error predictive discrepancy function, so the summary loss
functions are
\begin{align*}
  \LL_1(f, \gamma_1, X) &:= \sum_{i=1}^{n} [f(x_i) - \gamma_1(x_i)]^2,\\
  \LL_2(f, \gamma_2, X) &:= \sum_{i=1}^{n} [f(x_i) - \gamma_2(x_i)]^2
                          + [\lambda_1 \cdot J(h_1) +
                          \lambda_2 \cdot J(h_2)],
\end{align*}
with $J(h_j) = \int h_j''(t)^2 \dee t$, $j=1, 2$ is the complexity
penalty in the additive summary, and enforces smoothness in the
univariate functions $h_1$ and $h_2$.  The tuning parameters
$\lambda_1$ and $\lambda_2$ control the level of smoothness.  We do
not add a penalty for the linear summary.  The point estimates for
these summaries are found by minimizing posterior expected loss,
\begin{align}
  \hat{\gamma}_1(x) &= \arg \min_{\gamma_1 \in \Gamma_1}
                   \sum_{i=1}^{n}
                   [\hat f(x_i) - \gamma_1(x_i)]^2, \label{eq:point-est-lin}\\
  \hat{\gamma}_2(x) &= \arg \min_{\gamma_2 \in \Gamma_2}
                   \sum_{i=1}^{n} [\hat f(x_i) - \gamma_2(x_i)]^2 +
                   [\lambda_1 \cdot J(h_1) +
                          \lambda_2 \cdot J(h_2)].\label{eq:point-est-add}
\end{align}

The point estimate for the linear summary can be found via an ordinary
least squares fit to the vector of fitted values
$\hat{\mathbf{f}} = \{\hat f(x_i)\}_{i=1}^n$, i.e.
$[\hat \beta_1, \hat \beta_2]^\trans = (X^\trans X)^{-1}X^\trans
\hat{\f}$.  We find the projected posterior for the linear summary
using posterior draws of $\{f^{k}\}_{k=1}^M$, so that one draw from
the projected posterior is
$[\beta_1^{(k)}, \beta_2^{(k)}]^\trans = \arg \min_{\gamma_1 \in
  \Gamma_1}\LL_1(f^{(k)}, \gamma_1, X) = (X^\trans X)^{-1}X^\trans
{\f}^{(k)}$. 

The functions $h_1$ and $h_2$ for the additive summary are represented
by thin plate regression splines \citep[TPRS;][]{Wood2001} each with a basis
dimension of 10, with the identifying constraint
$\sum_{i=1}^n h_j(x_{ij})=0$ for all $j$.  Each function $h_j$ is
represented by the linear basis expansion,
$$h_j(x_j) = \sum_{m=1}^{M_j} \delta_{jm}\eta_{jm}(x_{j}) =
\sum_{m=1}^{M_j} \delta_{jm}z_{jm}$$ where the $\eta_{jm}$ are the
basis functions, and each function has $M_1=M_2=9$ basis terms.  The
entire vector of output from the additive model is given by
$\gamma(X) = \alpha + Z\delta$, where the $i$\textsuperscript{th} row
of the matrix $Z$ represents the linear basis expansion of $x_i$,
$z_i = (\{\eta_{1m}\}_m^{M_1}, \{\eta_{2m}\}_m^{M_2})$, and
$\delta = (\{\delta_{1m}\}_m^{M_1}, \{\delta_{2m}\}_m^{M_2})$ is the
concatenation of the basis weights.  The point estimate for the
additive summary \eqref{eq:point-est-add} is found by estimating
$\delta$ iteratively reweighted least squares, with the tuning
parameters $\lambda_1$ and $\lambda_2$ selected by minimizing the
generalized cross validation score on the values of $\hat f(x_i)$.  In
our implementation, we use the default settings of the \texttt{gam}
function in the \texttt{mgcv} package in \textsf{R}.  For details on
the form of the basis functions and how the model is fit, see
\cite{Wood2001,Wood2017}.  The particular choice of basis expansion is
not of main concern here, and any suitable basis will do.

In the end, the vector fitted values for the point estimate additive
summary \eqref{eq:point-est-add} can be represented by a linear
smoothing of the posterior mean fitted values from $f$, i.e.
$\hat{\gamma_2}(x) = P \hat{\mathbf{f}}$, with the influence matrix
calculated by $P = ZVZ^\trans$ where $V$ being the frequentist
covariance matrix of the estimates $\hat \delta$.  In fact, the fitted
values evaluated for each of the additive functions are the result of
a linear smoother, i.e.  $\hat h_j(X_j) = P_j \hat{\mathbf{f}}$, where
$P_j$ is the subset of rows of the projection matrix $P$ corresponding
to the basis expansion for the $j$th term.  This readily provides a
way to approximate the projected posterior for each smooth function in
the additive summary using posterior draws of original fitted values
$\f^{(k)}$.  A single posterior draw from the projected posterior is
calculated by $h_j^{(k)}(X_j) = P_j \f^{(k)}$.

\begin{figure}[!htb]
  \centering
  \includegraphics[width=0.9\textwidth]{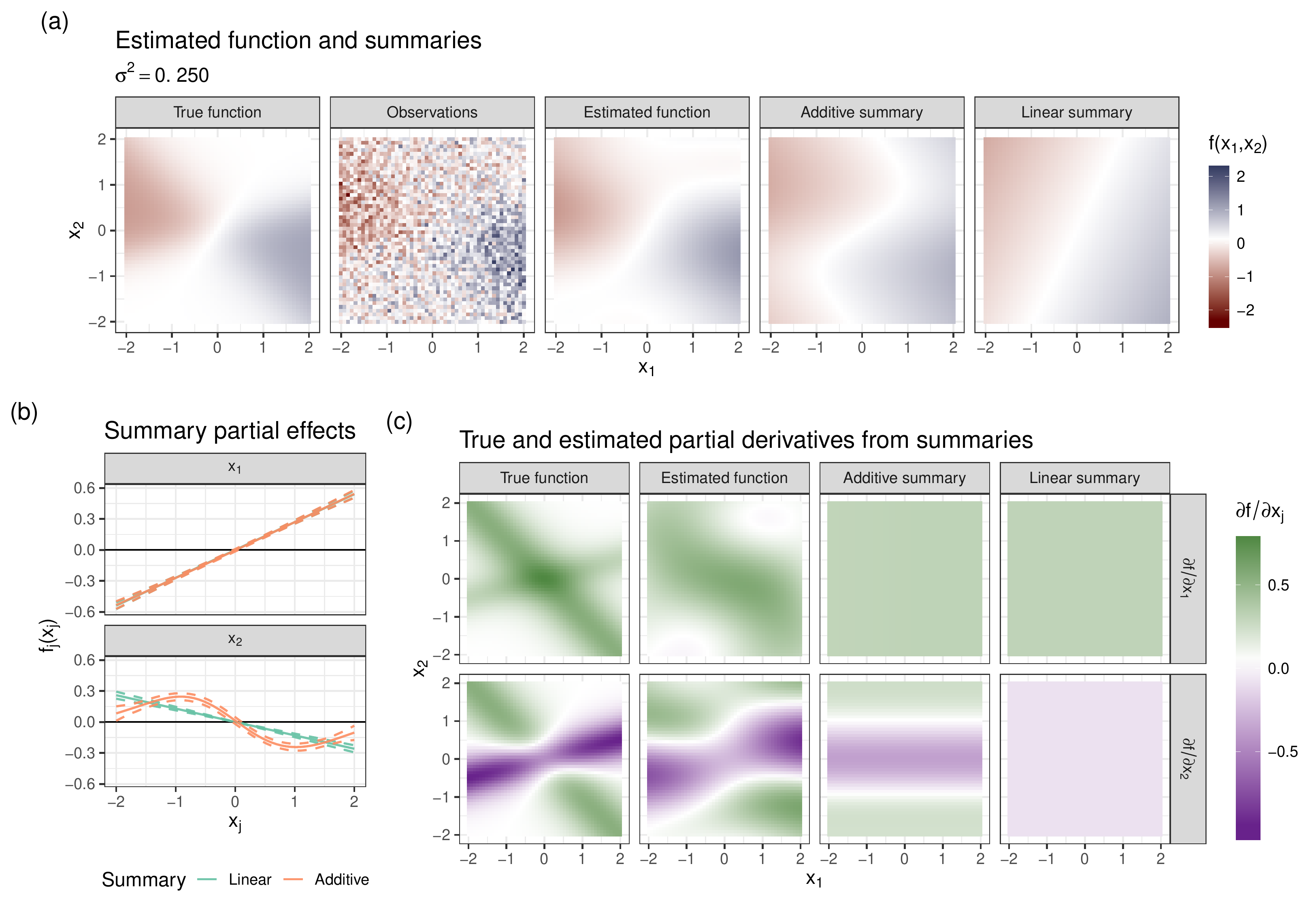}
  \caption{\label{fig:sigmoid-example} Result of simulated toy example
    summarizing a nonparametric bivariate function.  Panel (a) shows
    the nonparametric estimate of the true regression function, along
    with the regression surface resulting from the additive and linear
    summaries.  For for the linear summary, $R^2_{\gamma_1} = 75.9\%$
    and $\phi_{\gamma_1} = 7.6\%$ while for the additive summary,
    $R^2_{\gamma_2} = 82.4\%$ and $\phi_{\gamma_2} = 5.7\%$.  Panel
    (b) shows the estimated partial effects from the summary models.
    These partial effects estimate an average of the partial
    derivative of each covariate, assuming that it does not depend on
    the other covariate.  Partial derivatives of the true and
    estimated functions, and from the summaries, are shown in panel
    (c).  }
\end{figure}

Figure~\ref{fig:sigmoid-example} shows the resulting summaries.  Panel
(a) shows the true regression function and the observations, and
compares them to the estimated regression function from the GP model,
and to the bivariate surface resulting from the summary functions.
These summaries have differing degrees of fidelity in capturing
predictive trends in the original model; for the linear summary,
$R^2_{\gamma_1} = 75.9\%$ and $\phi_{\gamma_1} = 7.6\%$ while for the
additive summary, $R^2_{\gamma_2} = 82.4\%$ and
$\phi_{\gamma_2} = 5.7\%$.  Panel (b) shows the estimated partial
effects of each summary, along with 95\% credible bands from the
projected posterior.

These summaries estimate of the partial effect of each covariate.
Equivalently, they approximate the partial derivative of the true
regression function, with the (incorrect) simplifying assumption that
the partial derivative is constant in the other covariate (or constant
everywhere in the case of the linear summary).  Panel (c) shows the
partial derivatives with respect to $x_1$ and $x_2$ of the true,
estimated, and summary functions as a bivariate function of
$(x_1, x_2)$.  From this we can see that the summaries present
distinct ways of averaging the partial derivative from the estimated
regression function in a way that is readily presentable and
interpretable as partial effects in Panel (b).  We also quantify how
representative these summaries are of the original model with the
diagnostic measures.


\subsection{\textcolor{black}{Interaction detection in the presence of collinearity}}

\label{sec:illustr-toy-exampl}

The previous example showed how our method of posterior summarization
can communicate average partial effects.  Now we present a simulated
example with a more complex data structure to show how we can detect
significant interactions within a regression model, even in the
presence of collinear noise covariates, by using the iterative summary
search heuristics outlines in Section~\ref{sec:nonp-regr}.  We
consider data arising from the following:
\begin{align}\label{eq:new-sim-model}
  \begin{split}
    y
    & = f(x) + \varepsilon,
    \quad \varepsilon \sim \mathcal{N}(0, \sigma^2= 0.50)  \\
    f(x) & = 1 / \{1 + \exp(-2 x_1 x_2)\} + (x_3/3)^3.
  \end{split}
\end{align}
The true data generating process $f$ in Eq.~\eqref{eq:new-sim-model}
involves 3 variables $(x_1, x_2, x_3)$, with one interaction between
$x_1$ and $x_2$.  For this simulated example, we also observe 3 noise
covariates $(x_4, x_5, x_6)$ which are not involved in the data
generating mechanism, but are still correlated with the other
features.  To induce correlation between the features, we generate the
vector of covariates $[x_1, \ldots, x_6]^\trans$ from the multivariate
Gaussian distribution:
\begin{align*}
  [x_1, \ldots, x_6]^\trans
                 & \sim \mathcal{N}_6(0, \Sigma)    \\
  \Sigma
                 & =
    \begin{bmatrix}
      1          & 0 & 0.5 & \rho & \rho^2 & \rho^3 \\
      \vdots     & 1 & 0.5 & \rho & \rho^2 & \rho^3 \\
                 &   & 1   & \rho & \rho^2 & \rho^3 \\
                 &   &     & 1    & \rho   & \rho^2 \\
                 &   &     &      & 1      & \rho   \\
          \ldots &   &     &      & \ldots & 1  
    \end{bmatrix}
\end{align*}
with $\rho = 0.5$.  This reflects weak to moderate correlation between
the signal and noise covariates.  The two interactive features are
$(x_1, x_2)$ are uncorrelated, but both are correlated with the
remaining signal feature $x_3$.  We generate $n=400$ covariate vectors
of $[x_1, \ldots, x_6]^\trans$ from this Gaussian distribution to
create the design matrix $X$, and generate observations $Y$ using the
process in Eq.~\eqref{eq:new-sim-model}.  As in the previous example,
we estimate the regression function $f$ using a Gaussian process with
a squared exponential kernel and assign Jeffreys' prior to $\sigma^2$.
We obtain $1000$ draws from the posterior for $\mathbf{f}$.

\begin{figure}[ht!] \centering
  \includegraphics[width=0.7\textwidth]{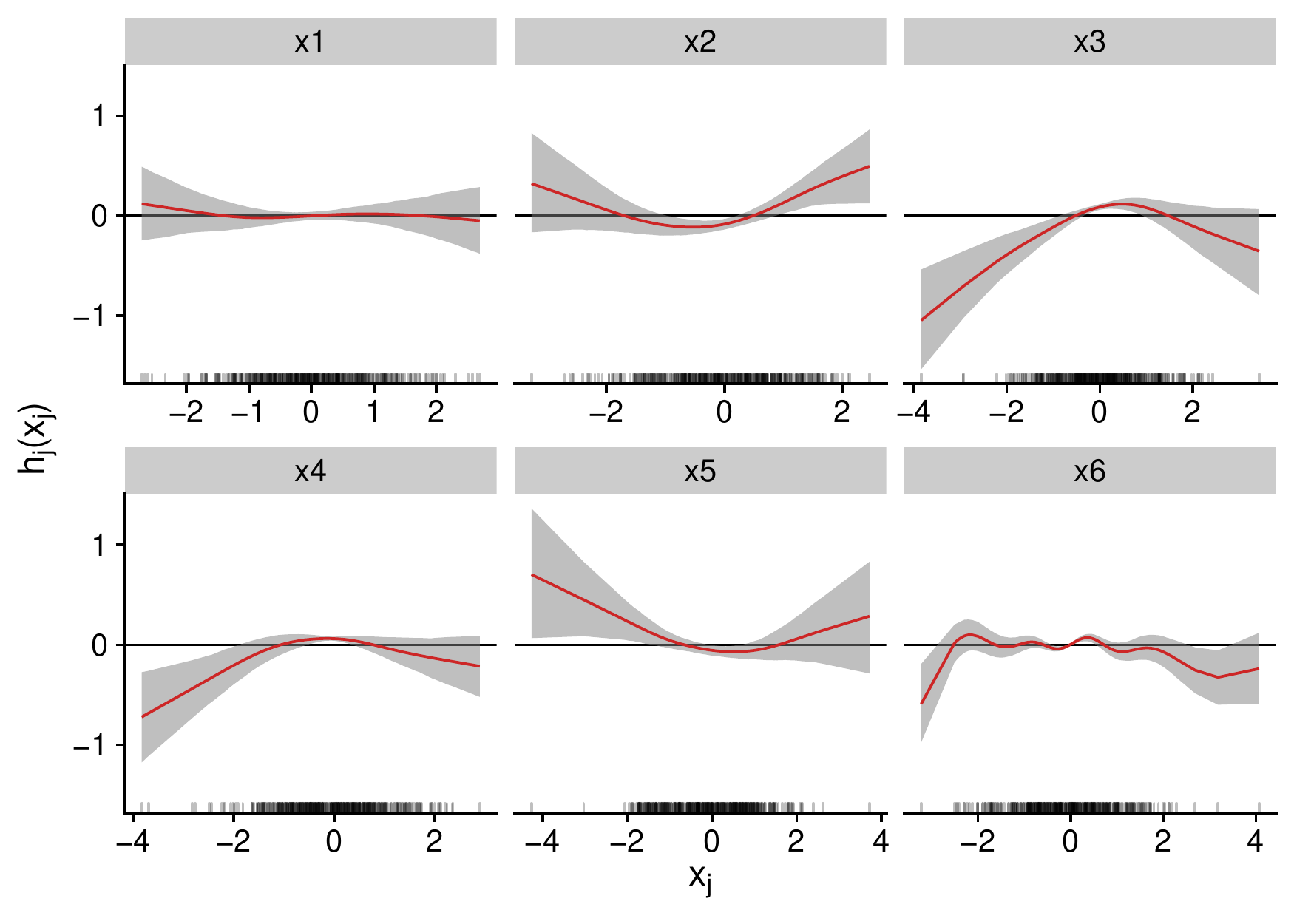}
  \caption{\label{fig:gam1-plot} Purely additive summary of
    toy example in
    Section~\ref{sec:illustr-toy-exampl}. 
  }
\end{figure}

To describe the predictive trends detected in the posterior for the
model $f$, we first construct an additive summary of the form
$\gamma_1(x) = \alpha_1 + \sum_{j = 1}^{6} h_j(x_j)$.  This summary
and its projected posterior are calculated in the same way as in the
previous example, and these are presented in
Figure~\ref{fig:gam1-plot}.  To see how well this summary describes
the fitted model, we first we consider the predictive variance
explained by the summary, which is $R^2_{\gamma_1} = 61\%$ in this
case, suggesting that there is considerable amount of variation in the
posterior for $f$ that is not being captured by this summary.  Second,
because this summary assumes additivity in the covariates, we can
detect significant sources of unexplained variation due to
interactions by constructing a single decision tree regressing the
summarization residuals $\hat f(x_i) - \hat \gamma(x_i)$ on all the
observed covariates.  This tree is shown in the supplement, and the
pairs of covariates in neighboring nodes of this tree suggest
prospective significant interactions detected by the model which
should be accounted for.

To create a more faithful summary of the fitted function, we consider
summaries which allow for a two-way interaction between two of the
covariates, leaving the summary function additive in all the other
covariates.  That is, the summary class is $\Gamma_{kl}$, the set
functions of the form
$\gamma_{kl}(x) = h_{kl}(x_k, x_l) + \sum_{j \notin \{k,l\}}
h_j(x_j)$.  This summary function is partially additive, allowing for
a single two-way interaction via the bivariate function
$h_{kl}(x_k, x_l)$, represented as a two-dimensional thin plate
regression spline with basis dimension 30.  The summary loss function
in this case is
\begin{align*}
  \LL_{kl}(f, \gamma_{kl}, X) &:= \sum_{i=1}^{n} [f(x_i) -
                                \gamma_{kl}(x_i)]^2
                                + 
                                \left[
                                \lambda_{kl} J_{kl}(h_{kl})
                                + \sum_{j\notin\{k,l\}}^{} \lambda_j
                                J(h_j)
                                \right]. 
\end{align*}
with the $J_j$ penalties defined before, and $J_{kl}$ now enforcing
smoothness in the bivariate function.  For details of higher-order
TPRS penalty functions we again refer the reader to
\cite{Wood2001,Wood2017}.  The point estimate summary and projected
posterior for $\gamma_{kl}$ are found analogously to those of the
purely additive summary.

The relevant question is the choice of which interaction $(x_k, x_l)$
to add, which should be the most significant interaction present in
the fitted model.  We could narrow our attention to considering in
interactions between the pairs of covariates detected by the suggested
by the summary residual regression tree, but for the sake of
completion, we fit the summary for each possible pair of covariates.
Whichever interaction pair leads to the greatest increase in
$R^2_\gamma$ should indicate which interaction is most significant in
the model.

\begin{figure}[ht!]
  \centering
  \includegraphics[width=0.6\textwidth]{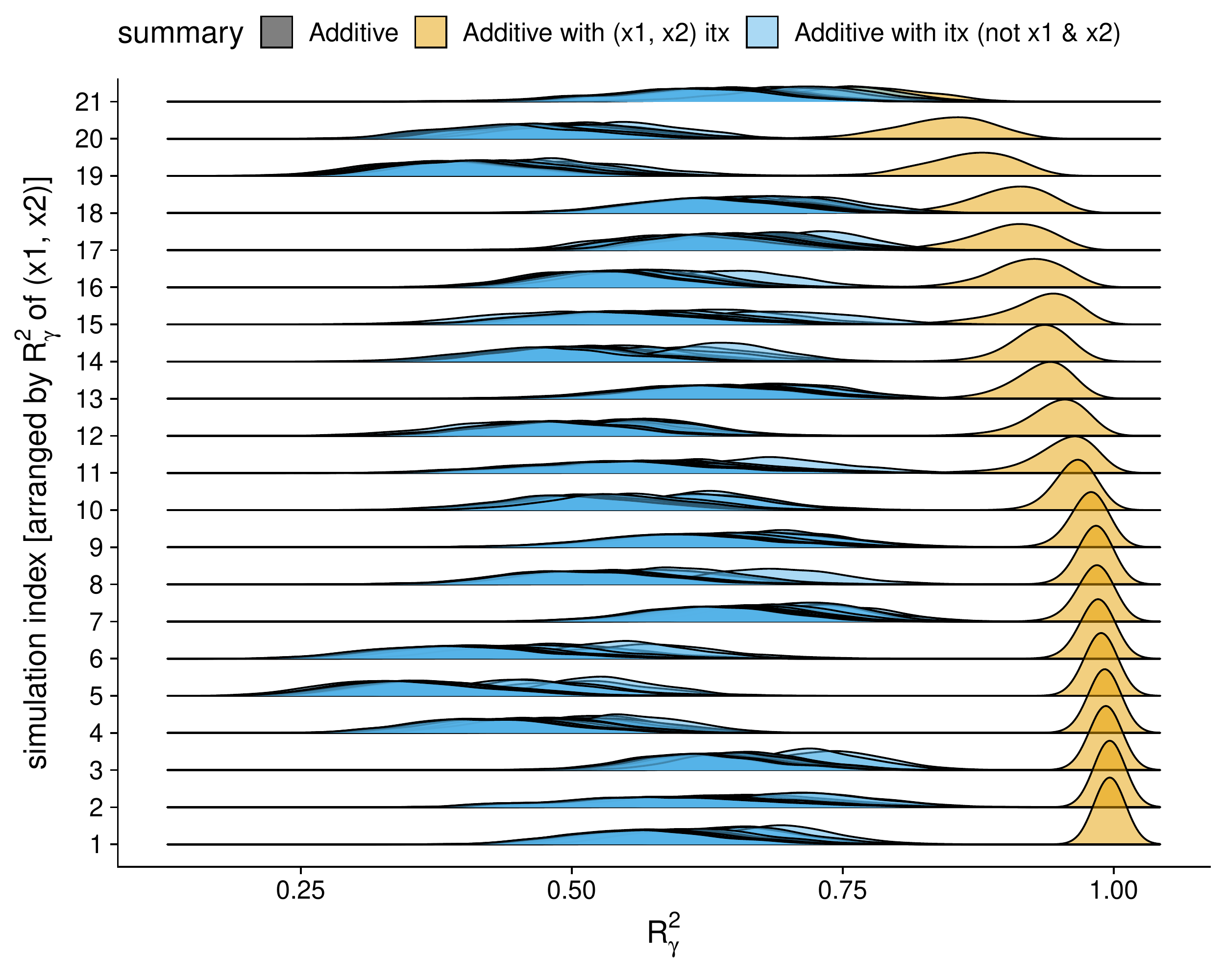}
  \caption{\label{fig:rsq-ridges} Variation explained by different
    summaries of the Gaussian process regression model in
    Section~\ref{sec:illustr-toy-exampl}, including purely additive
    summary (``Additive''), partially additive summary with an
    $(x_1, x_2)$ interaction (``Additive with $(x_1, x_2)$ itx''), and
    partially additive summaries with two-way interaction which is not
    \emph{not} between $(x_1, x_2)$ (``Additive with itx
    $(\text{not } x_1 \& x_2)$ itx'').  The bottom row corresponds to the
    initial generated dataset; the other rows correspond to twenty
    replications of the same simulated example.  Allowing for the
    $(x_1, x_2)$ interaction in a partially additive summary routinely
    gives the compared to the purely additive summary.  }
\end{figure}

\begin{figure}[ht!]
  \centering
  \includegraphics[width=0.475\textwidth]{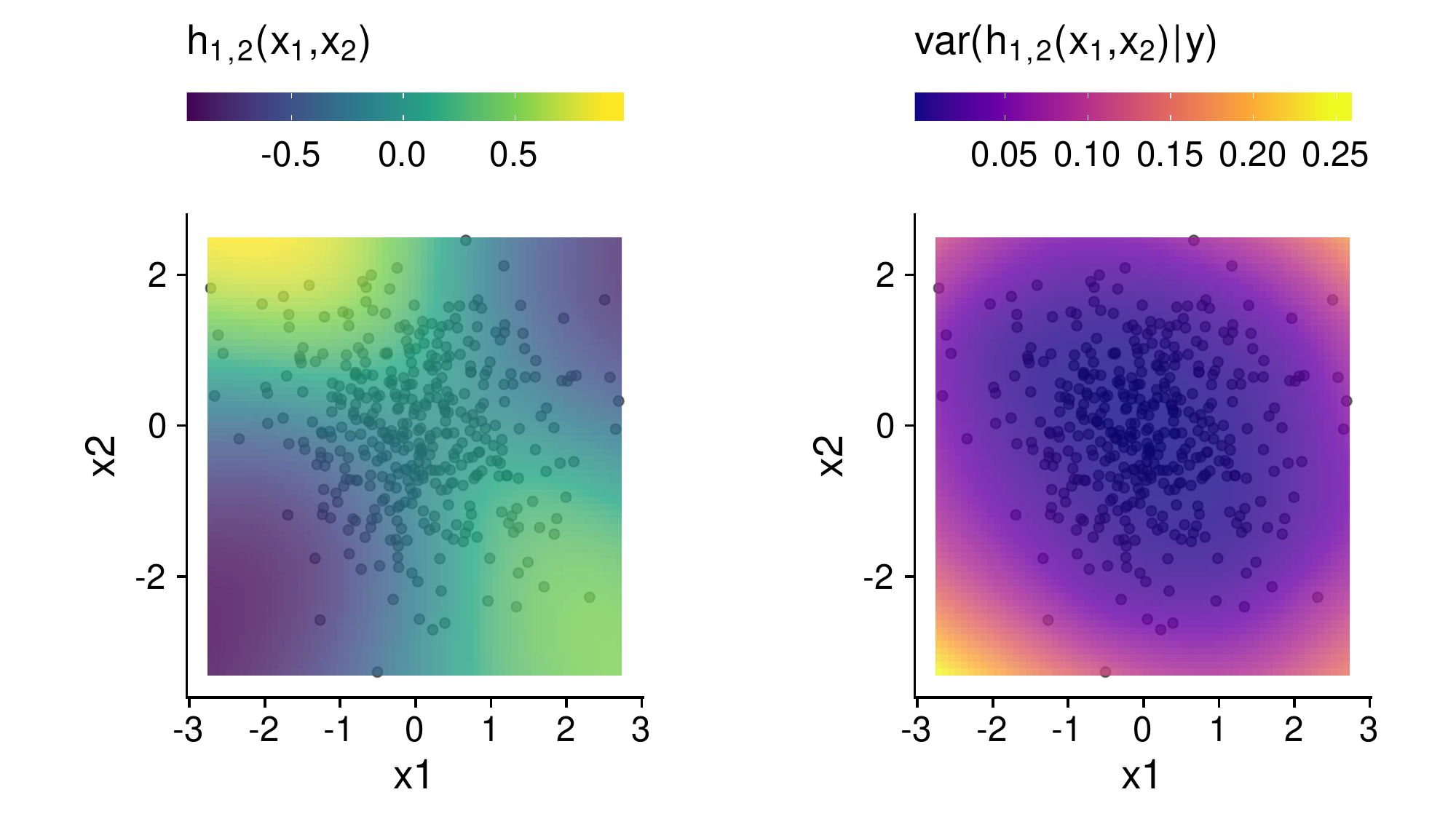}
  \includegraphics[width=0.475\textwidth]{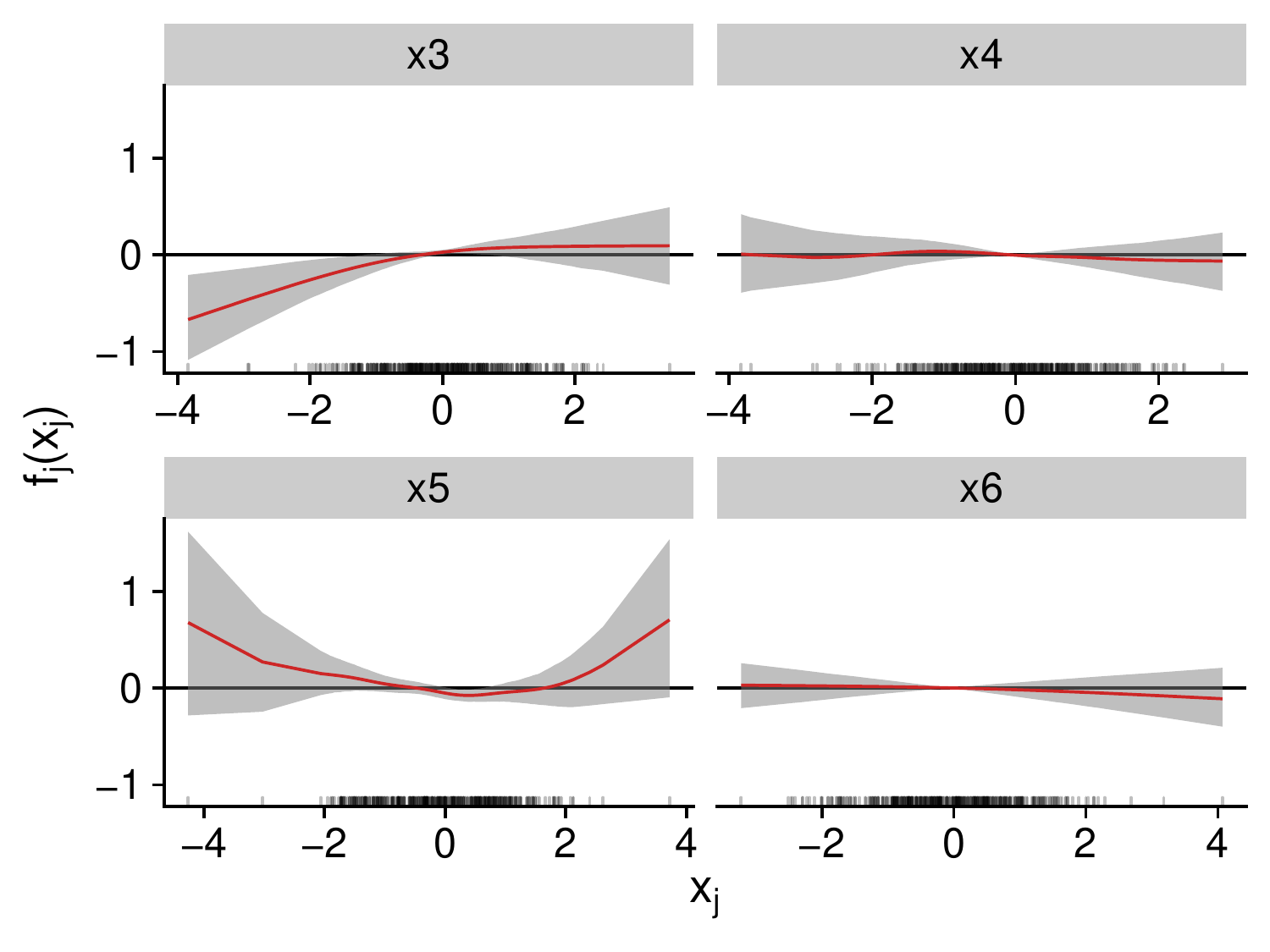}
  \caption{\label{fig:gam2-plot} Partially additive summary with a
    bivariate interaction for $(x_1, x_2)$ for the simulation example
    in Section~\ref{sec:illustr-toy-exampl}}
\end{figure}

The $R^2_\gamma$ values for these partially additive summaries are
show in the bottom row of Figure~\ref{fig:rsq-ridges}.  The
interaction between $x_1$ and $x_2$ gives by far the biggest gain in
predictive variance explained as measured by $R^2_\gamma$ compared to
every other possible two-way interaction, with this metric rising to
about $R^2_{\gamma_{12}}=96\%$.  This confirms that the interaction
for $(x_1, x_2)$ is the most important one detected by the GP model
for $f$.  The resulting partially additive model summary with an
$(x_1, x_2)$ interaction for this simulation is shown in
Figure~\ref{fig:gam2-plot}.

This pattern is routinely detected across multiple replications of
this simulation example; when replicating this simulation up to 1000
times, the $(x_1, x_2)$ interaction gives the largest boost in
$R^2_\gamma$ at a rate of 98.9\% (SE = 0.3\%).  The $R^2_\gamma$
values for twenty of these additional replications are given in
Figure~\ref{fig:rsq-ridges}, with the comprehensive set of results
presented in the supplement. This suggests that our method is able to
recapitulate the interaction which is present in the true data
generating process when summarizing a Gaussian process regression,
even in the presence of correlated noise variables.  Note that this
results hinges on the specified regression model being able to
accurately capture the interaction between the covariates.

\section{Application to California housing data}
\label{sec:exampl-calif-hous}

Here we demonstrate our approach using data from the 2011 American
Community Survey on housing prices in California at the census tract
level.  We regress census tract log-median house value on log-median
household income, log-population, median number of rooms per unit,
longitude, and latitude, using a Gaussian process regression model.
The full model is given by
\begin{align*}
  \begin{split}
    (y_i \mid f, \sigma^2) = f(x_i) + \varepsilon_i, \quad
    \varepsilon_i \sim \N(0, \sigma^2) \\
  f \sim \text{GP}(0, k(\cdot, \cdot)), \quad p(\sigma^2) \propto \sigma^{-2}
  \end{split}
\end{align*}
where $y_i$ is the log-median house value and $x_i$ is the covariates
for census tract $i$.  We use a combination of squared exponential
kernel and the linear covariance kernel,
\begin{align*}
  k(x_i, x_{i'}) = \tau^2 \cdot \exp
  \left(
  - \sum_{j=1}^p   [x_{ij} - x_{i'j}]^2 / v_j
  \right) +
  \sum_{j=1}^{p} a_j x_{ij} x_{i'j}
\end{align*}
for the $p=5$ predictors.  Empirical Bayes estimates for
$\hat{\sigma}^2$, $\hat\tau^2$, $\{\hat{v}_j\}$, and $\{\hat a_j\}$
were found using maximum marginal likelihood estimation.  We obtained
$1000$ posterior draws for $\sigma^2$ and $f$ using MCMC after fixing
the GP hyperparameters to the estimated values $\hat\tau^2$,
$\{\hat{v}_j\}$, and $\{\hat a_j\}$.  Denote by $\mathbf{f}^{(k)}$ the
vector of fitted values at all covariate locations in the dataset $X$
for the $k$\textsuperscript{th} Monte Carlo posterior draw of $f$, for
$k = 1, \ldots, M=1000$.  The GP model can account for nonlinear and
interactive effects of covariates on housing prices, and because of
this flexibility, we achieve a good quality of fit as measured by the
usual coefficient of determination, $R^2 = 83\%$.

However, the output of the fitted GP model alone has little utility
for qualitatively understanding the influence of each covariate.  To
better understand the fit, we calculate several summaries for this
regression model, each representing different characterizations of the
relationship between the covariates and the output, as an illustration
of the iterative approach outlined in Section~\ref{sec:nonp-regr}.  We
leave out most of the technical details of calculating the summaries,
as they closely mirror those of the simulation examples in
Section~\ref{sec:simul-results-nonp}.

We first consider global summaries of model behavior, showing how the
class of summaries can be refined until it is deemed a satisfactory
representation of the original model's predictions, and also how this
process can reveal important interactive effects in the housing price
model.  Then we compute local summaries of model behavior,
investigating how determinants of housing prices differ
geographically.  We only consider linear summaries for explaining
local behavior, but demonstrate how adjusting the level of locality
detects heterogeneity in covariate importance between these local
areas.

\subsection{Global summary search}
\label{sec:glob-summ-search}

\subsubsection{Global linear summary}
\label{sec:glob-line-summ}

We start by creating a linear summary for the fitted model, perhaps
the simplest summary one could make of a nonparametric
regression.  The summary function has the form
$\gamma(x) = x^\trans \beta$.  The vector $\beta$ represents the
average partial effect of each covariate.  There is no penalty term
used here (imposing linearity is already a significant restriction),
but one could just as easily use a penalty term if a sparse linear
summary is desired.  The point estimate and projected posterior for
the linear summary are calculated in a parallel manner to those from
the simulation example in Section~\ref{sec:toy-exampl-nonadd}.

\begin{figure}[!htb]
  \centering
  \includegraphics[width=0.7\textwidth]{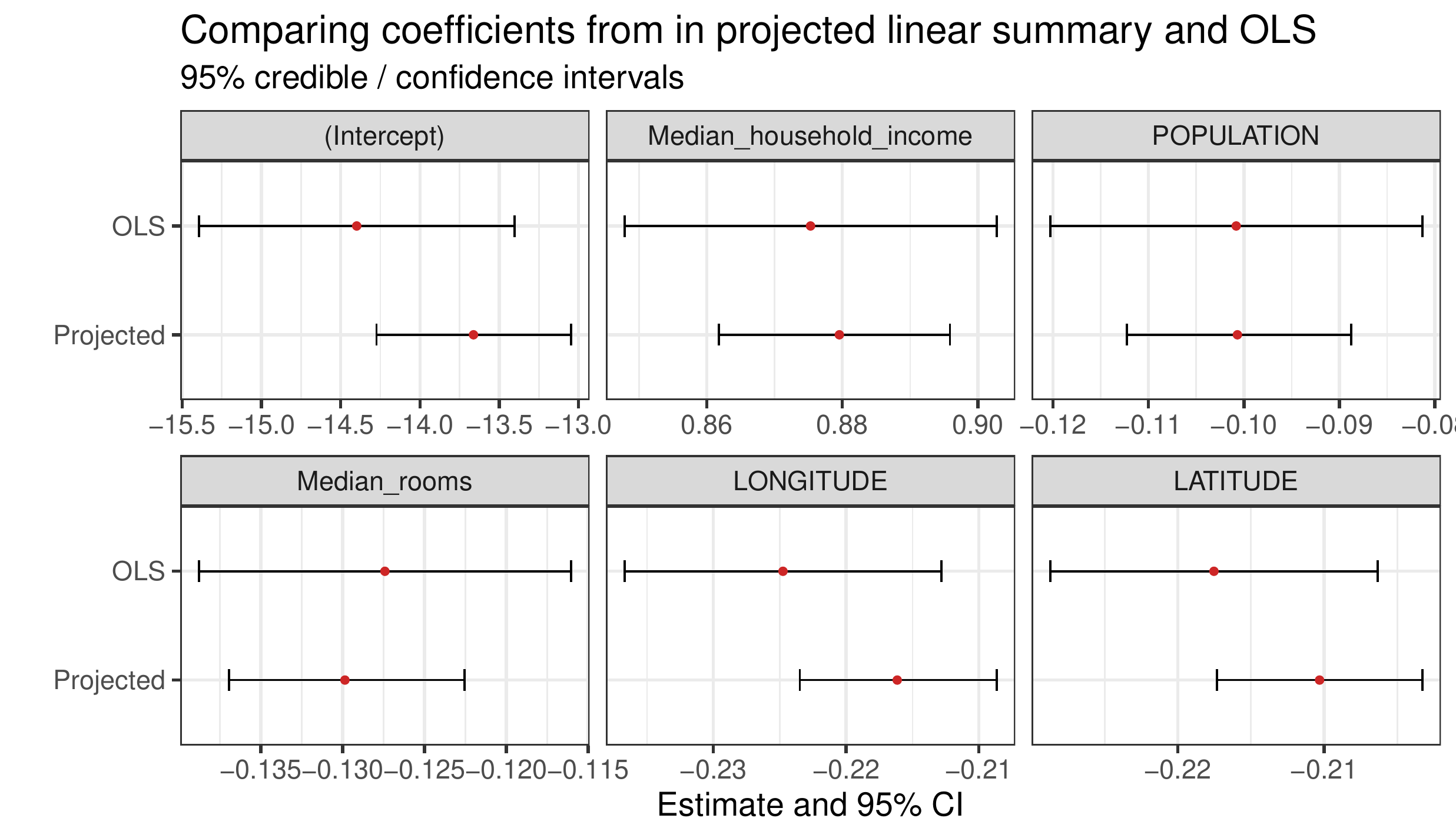}
  \caption{Comparison of projected linear summary of GP regression vs
    OLS regression, comparing results of linear model regressing $X$
    on $y$, with coefficient estimates and 95\% confidence intervals
    (for OLS regression), and 95\% projected credible intervals (for
    linear summary).  For this linear summary, $R^2_\gamma = 66\%$ and
    $\phi_\gamma = 57\%$.  While this suggests a poor quality of model
    summary, it still represents the best linear approximation to the
    regression surface.  Projected credible intervals are
    appropriately narrower than the confidence intervals from OLS, as
    this is a summarization of the full GP model rather than being
    considered the ``true'' model.  Point estimates are generally
    pulled toward zero as an result from the shrinkage effect of the
    GP.  }
  \label{fig:linear-summary}
\end{figure}

Figure~\ref{fig:linear-summary} shows the results of the projection
and, compared to the results of fitting an ordinary least squares
(OLS) regression of $y$ on $X$. On average, the projected credible
intervals for the coefficients in the linear summary are about 30\%
narrower than the 95\% confidence intervals from OLS. Also, point
estimates are generally closer to zero for the linear summary than for
OLS, likely due to a shrinkage effect from the GP prior.  In a sense,
this is precisely what we would expect to see.  The linear summary is
the best linear approximation to the fitted function $f$ from the GP,
without assuming that the response surface is actually linear.
Furthermore, the linear summary is a projection of the fitted values
from $f(x_i)$, which have lower variance than the observations used
for creating the OLS estimates.

\begin{figure}[!htb]
  \centering
  \includegraphics[width=0.6\textwidth]{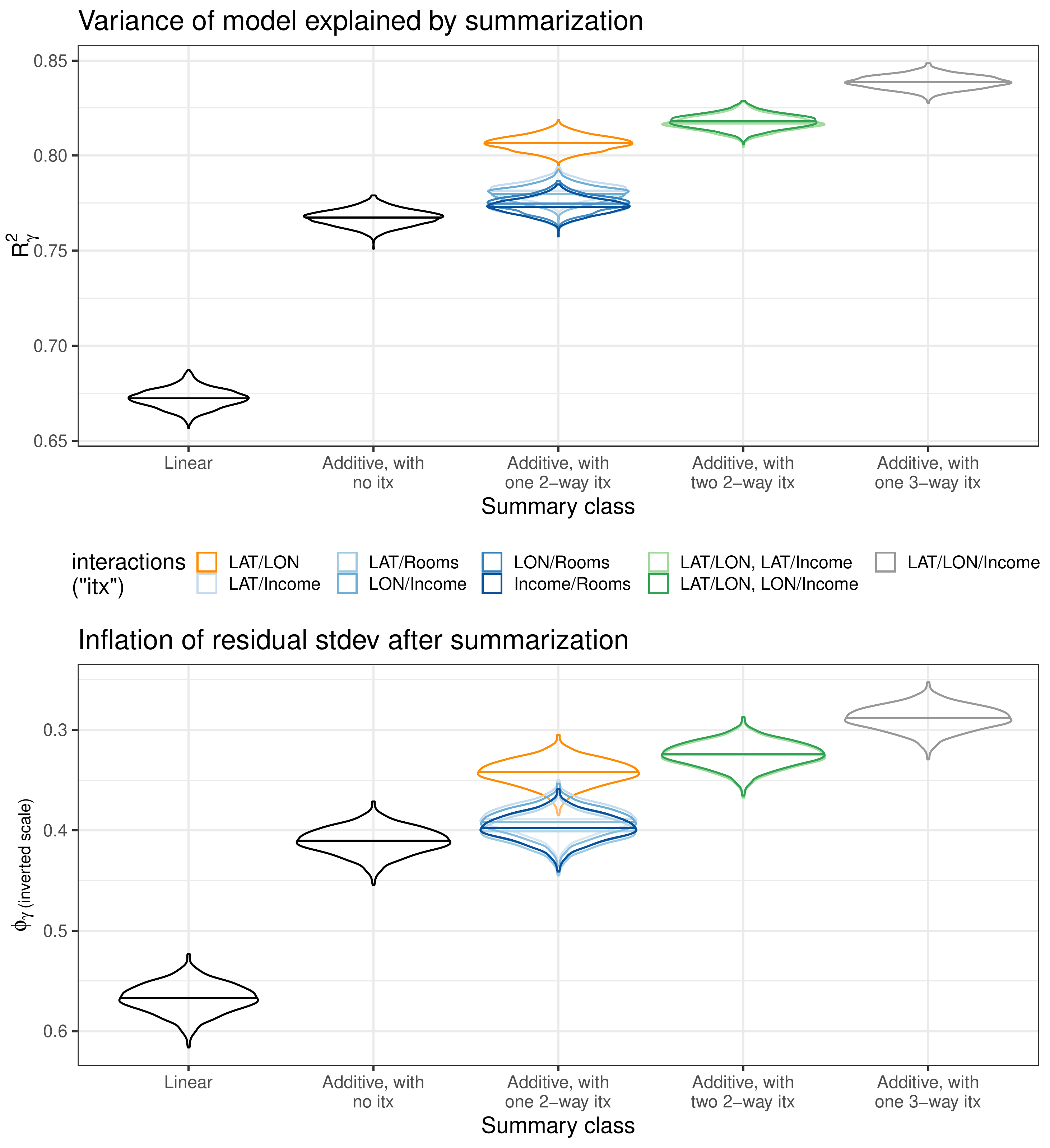} \\
  \caption{\label{fig:summary-plots} Diagnostics for quality of model
    summarization for increasingly complex summaries.  We start with a
    linear summary, then an additive summary (GAM), and then
    progressively adding two-way interactions and finally a three-way
    interaction to the GAM.  We consider interactions among different
    combinations of covariaties.  Horizontal lines within violin plots
    indicate posterior medians.  We choose to report the additive
    model summary with one two-way interaction for longitude and
    latitude, as this summary has a good degree of predictive
    explainability while still being interpretable and easily
    communicable.  This summary is presented in Figure
    \ref{fig:gam-2}.  \sw{Change $x$-axis to say ``purely additive''
      and ``partially additive''}}
\end{figure}

The diagnostics for this linear summary are shown in
Figure~\ref{fig:summary-plots}, along with those from several other
fitted summaries (which will be described later).  The linear summary
explains about $R^2_\gamma = 66\%$ of the variation in the predictive
model, and residual standard deviation is inflated by about
$\phi_\gamma = 57\%$.  This reflects a rather poor summary
representation of the model, and suggests that there is important
variation in the regression model that is being unaccounted for.
While this summary is indeed the best linear approximation to the
fitted regression model, we are evidently missing out on important
features of $f$.

\subsubsection{Global additive summary}

The requirement of linearity is rather limiting for summarizing the
fitted GP regression, so we remove this constraint and consider
instead the larger class of additive functions.  Instead of a
describing the partial effects of covariates on housing pricess
linearly, we now describe partial effects with smooth nonlinear
functions.  That is, the summary class $\Gamma$ comprises functions of
the form
\begin{align}\label{eq:gam}
  \gamma(x) &= \alpha + \sum_{j=1}^{5} h_j(x_{j}),
\end{align}
and, again as in the simulation example in
Section~\ref{sec:simul-results-nonp}, each function has a thin plate
regression spline representation with basis dimension 10.

\begin{figure}[!htb]
  \centering
  \includegraphics[width=0.65\textwidth]{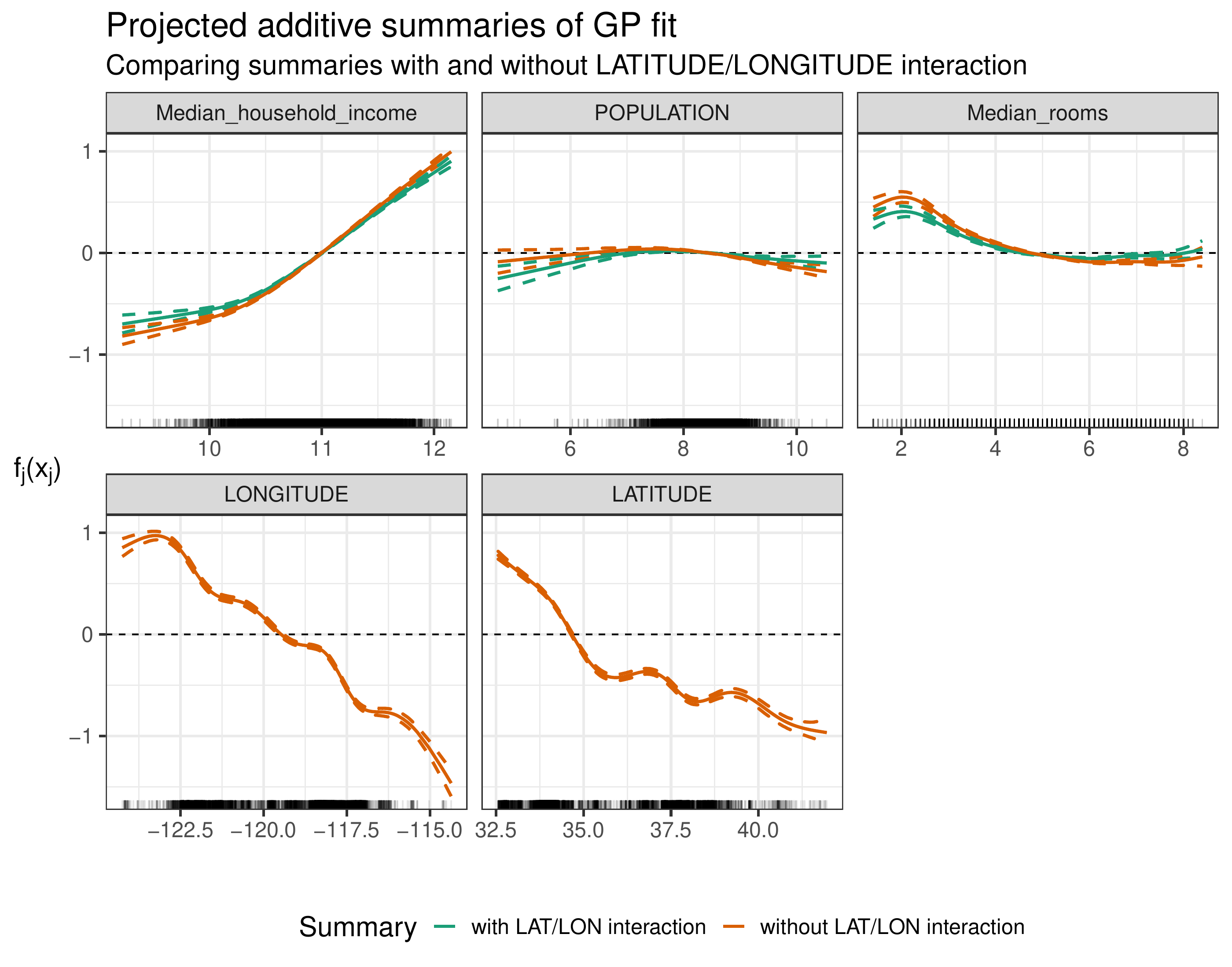}
  \includegraphics[width=0.34\textwidth]{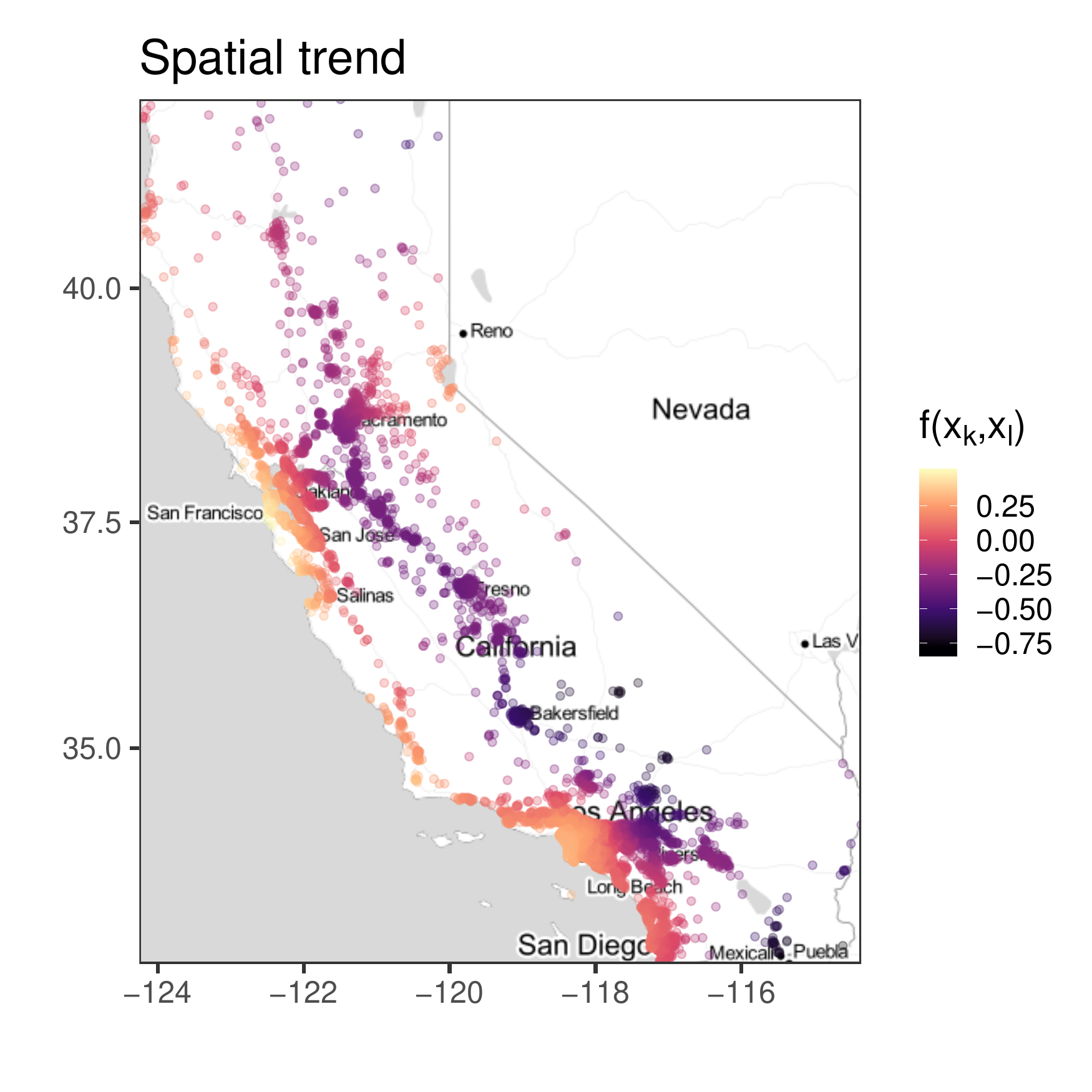}
  \caption{\label{fig:gam-2} Additive summary of GP regression model
    before and after adding spatial interaction.  }
\end{figure}

The point estimate and 95\% credible bands for this additive summary
are represented by the orange lines in
Figure~\ref{fig:gam-2}. Diagnostics for this summary are shown in the
second column of Figure~\ref{fig:summary-plots}.  Compared to the
linear summary, the additive summary \eqref{eq:gam} represents a
significant gain in predictive explainability as measured by both
$R^2_\gamma$, rising from $66\%$ to $76\%$, and $\phi_\gamma$,
dropping from $57\%$ to $40\%$.

Still, the assumption of additivity is quite a strong one for
summarizing the fitted GP regression.  There may be significant
underlying interactions in the original model which we are missing
here.  To investigate this possibility, we fit a regression tree to
the summary residuals $\hat f(x) - \hat \gamma(x)$ truncated to a
maximum depth of four for ease of presentation (shown in the
supplement).  The tree detects a high degree of heterogeneity in the
summary residuals, so we next consider allowing for interactions in
our summary.


\paragraph{Interaction search}

Analysis of the summary residuals from the additive summary suggests
that we should refine the summary to allow for some low-level
interaction among the covariates.  Specifically, it appears that
longitude and latitude have the most important interactive effect, as
these covariates appear highest in the summary residual regression
tree.  For the sake of completeness, we will also consider
interactions involving median rooms and log-median household income,
as these covariates also appear in the regression tree (even though
very few data points fall into the nodes corresponding to these
covariates; we do exclude the log-population covariate from
consideration, as the node containing this variable contains a
vanishingly small proportion of data points).

We will initially consider adding a single two-way interaction to the
summary, using every possible pairing of these four covariates.  Then we will move
along a path of increasing summary complexity, adding a second
two-interaction, and finally considering an additive summary with a
three-way interaction.

Including a single two-way interaction, $\Gamma$ is now to the set
of partially additive~functions
\begin{align} 
  \label{eq:gam-itx}
  \gamma(x) &= \alpha + h_{kl}(x_k, x_l) + \sum_{j \notin \{ k,l \}} h_j(x_j),
\end{align}
where $h_{kl}(x_k, x_l)$ is a two-dimensional smooth function for the
$(x_k, x_l)$ interaction, constructed using a two-dimensional thin
plate regression spline with basis dimension
30. 

Figure~\ref{fig:summary-plots} contains the summary diagnostics for
all the considered configurations of the partially additive summary
\eqref{eq:gam-itx} using the specified covariates.  As suggested by
the summary residual regression tree, the additive summary interacting
longitude and latitude marks the best improvement by far in predictive
explainability, marking a rise in $R^2_\gamma$ from $76\%$ to $81\%$
and a fall in $\phi_\gamma$ from $40\%$ to $35\%$ as compared to the
original (non-additive) summary. That this is the most significant
interactive effect is not surprising, as geography likely plays a
large role in explaining housing prices in a way that is not separable
by latitude and longitude.  The fitted summary, accompanied by 95\%
credible bands, is shown in Figure~\ref{fig:gam-2} in comparison to
the previously fitted non-interactive additive summary.

Again, we look for the possibility of an important unaccounted
interactive effect by fitting a regression tree to the summary
residuals from this newly calculated summary (shown in the
supplement).
Longitude and latitude seem to remain the most significant source of
summary residual heterogeneity, possibly implying that the fitted
two-dimensional smooth function in the interactive additive summary
was oversmoothed.  However, we turn our attention now to possible
interactions between median household income and the spatial
covariates, which are implied by this second summary residual
regression tree.


We consider introducing a second two-way interaction in addition to
the longitude-latitude interaction.  That is, we consider two
summaries, including (i) a summary allowing for interactions for
longitude-latitude and longitude-income, and (ii) a summary allowing
interactions for longitude-latitude and latitude-income.  However,
neither of these summaries mark a significant improvement over the
summary with a single two-way interaction, demonstrated by the fact
that the posteriors for the summary diagnostics of these two models
overlap with that of the partially additive summary with only the
longitude-lattitude interaction, seen in
Figure~\ref{fig:summary-plots}.

The next step up in the progression of summary complexity is to
accommodate a three-way interaction for longitude-latitude-income
(i.e., a three-dimensional smooth).  Looking at the summary
diagnostics for this fitted summary, we do now notice a significant
gain in predictive explainability over the partially additive summary
with a single interaction for longitude-latitude.  But choosing this
summary model would require a large sacrifice in interpretability of
the summary for a relatively low gain in predictive ability.

Therefore, we conclude the summary model with one interaction between
latitude and longitude is most appropriate to report.  It has an
$R^2_\gamma$ value of about $81\%$ and a $\phi_\gamma$ value of about
$35\%$, which is considerable given the level of complexity which the
original GP regression model is able to accommodate.  Thus, we can
conclude that the trend in housing prices as explained by the
covariates is somewhat close to additive, with an important
interaction between longitude and latitude, although some more complex
features remain.

\subsection{Local linear summaries}
\label{sec:local-linear-summary}

To draw out some of these features, we consider local behavior of the
regression function $f$.  Previously we focused on global model
summaries, capturing how the model behaves on average across the
entire dataset.  However, one of the advantages of nonparametric
regression is that the model adapts to heterogeneity in the response
surface. That is, covariate importance is likely to be nonconstant
across the covariate space. This applies in our example; it is likely
true that determinants of housing prices vary geographically.

Given this feature, we now investigate the geographic variation in how
covariates influence housing prices.  We selected three metropolitan
areas in California for comparison.  These represent the southern,
central, and northern regions of the state, with these areas defined
by their encompassing counties: Greater Los Angeles (LA and Orange
Counties), Fresno (Fresno County), and the Bay Area (San Francisco and
San Mateo Counties).  We calculate local linear summaries at four
different resolutions: (i) one summary for each of the metropolitan
areas, (ii) one for each of the constituent counties for these
metropolitan ares, (iii) for several neighborhoods within one of these
counties, and (iv) for one specific census tract.  These local linear
summaries explain how the model makes predictions at these geographic
levels, and describe the relative importance of each covariate to each
area.

For each of these localities, we computed linear summaries of the
output of the fitted GP regression model using the following
procedure.  First, generate $\tilde{n}=1000$ new geographic locations
to represent newly generated census tracts by sampling uniformly
within these areas (in the case of the linear summary of the single
census tract, we fix the location at this one point and simply create
$\tilde{n}=1000$ copies).  Next, for each of these synthetic
geographic locations, generate values for the other covariates.  For
this step we calculated the empirical mean and covariance of the three
non-geographic covariates at the metropolitan area level, and drew
samples from the three-dimensional Gaussian distribution with these
parameters.  These two pieces collectively define the full set of
predictive locations $\tilde X$ for the locality under consideration.
Then, for each of these newly created data points, we obtain $M=1000$
MCMC posterior draws of the output of the fitted regression function,
and calculate the linear summary by projecting the fitted values from
the full model onto the column space of $\tilde X$, similar to the
process described in Section~\ref{sec:glob-line-summ} for the global
linear summary.

\begin{figure}[!htb]
  \centering
  \includegraphics[width=0.4\textwidth]{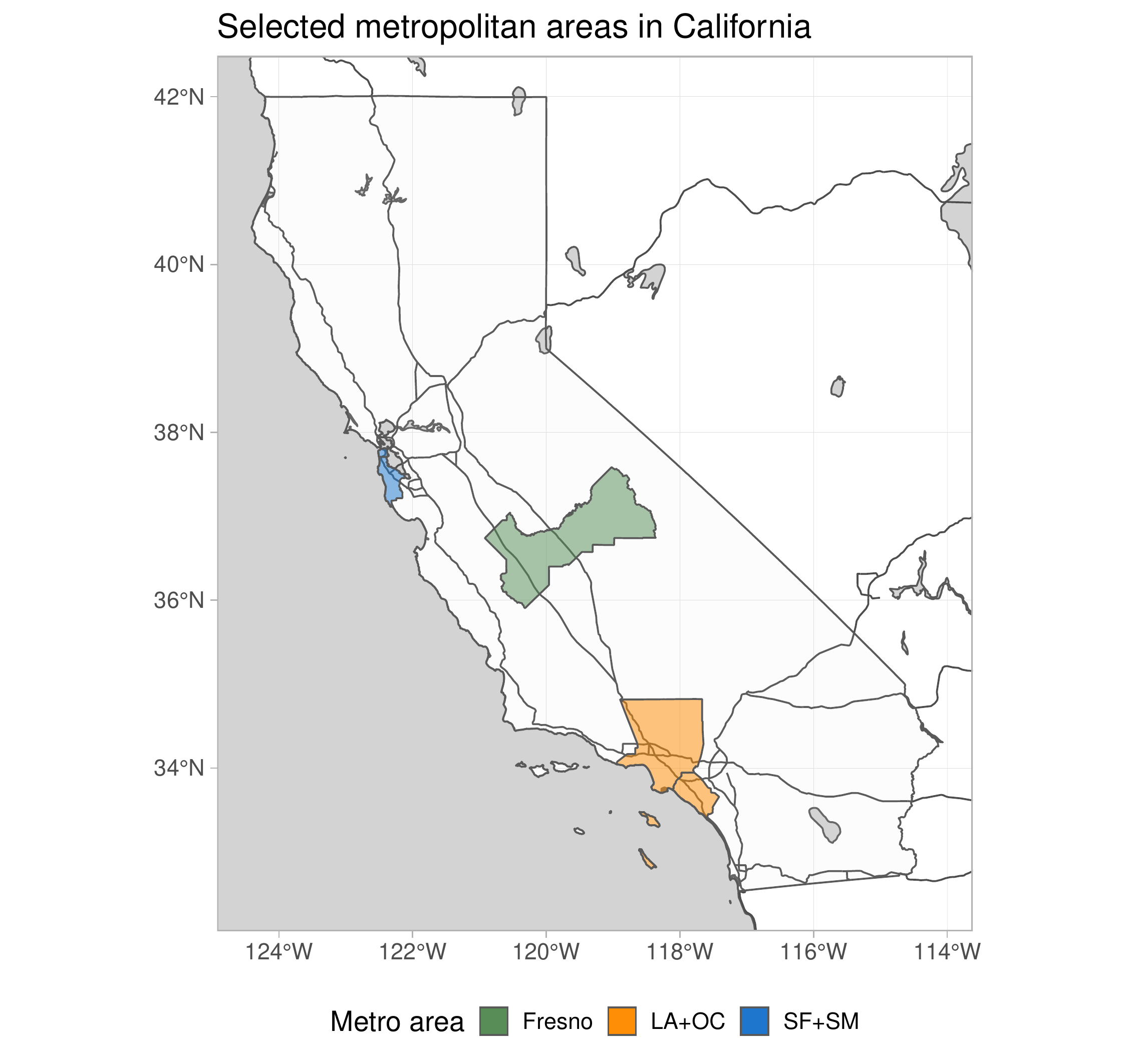}
  \includegraphics[width=0.5\textwidth]{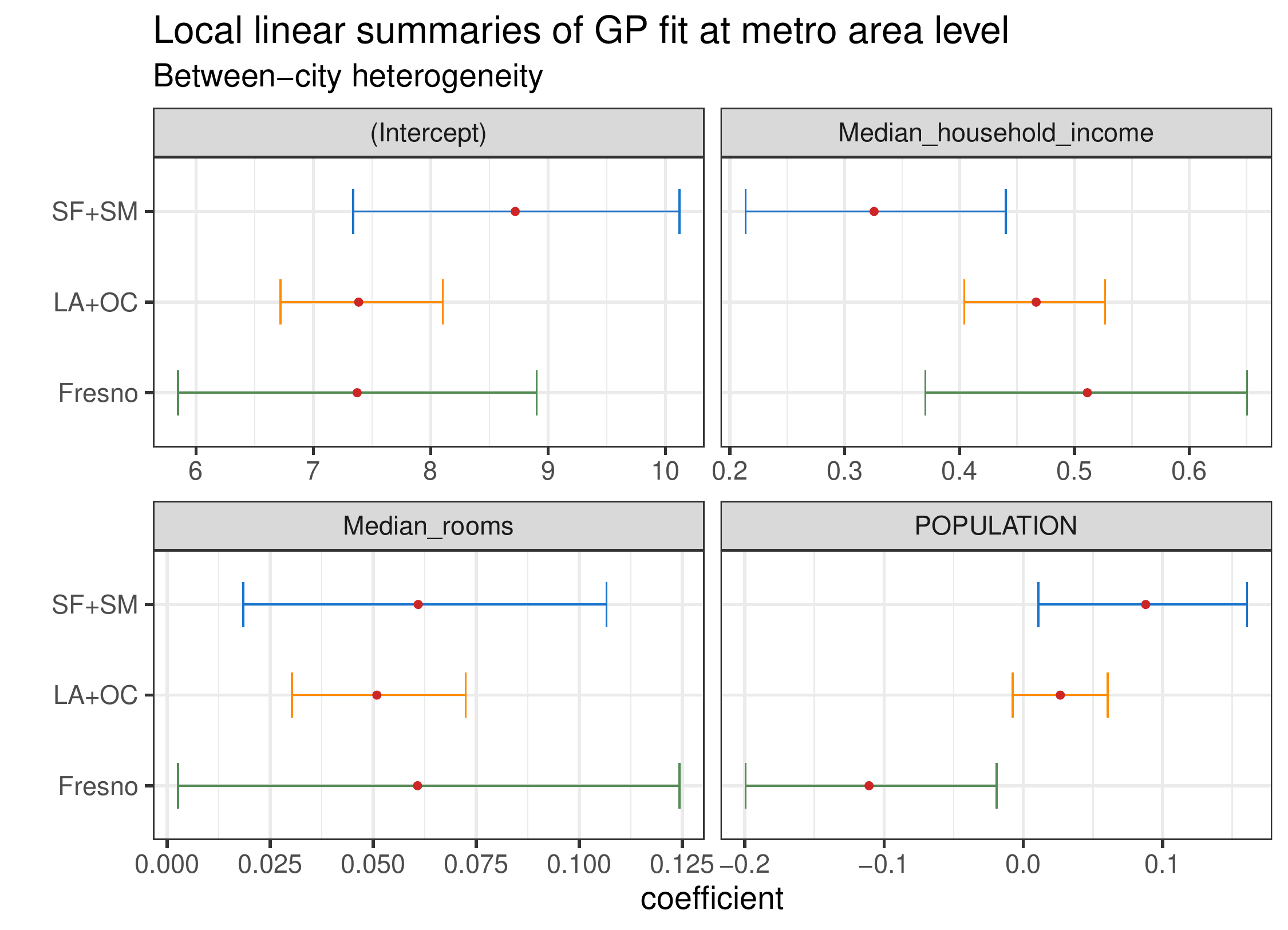}
  \caption{\label{fig:selected-counties} Local linear summaries
    produced for metropolitan areas in California.  Metropolitan areas
    are defined by their counties, and we selected the Bay Area (San
    Francisco and San Mateo Counties), Greater Los Angeles (Los
    Angeles and Orange Counties), and Fresno (Fresno County).
    Determinants of housing prices do vary quite notably, particularly
    for median income and population. }
\end{figure}

\begin{figure}[!htb]
  \centering
  \includegraphics[height=0.3\textheight]{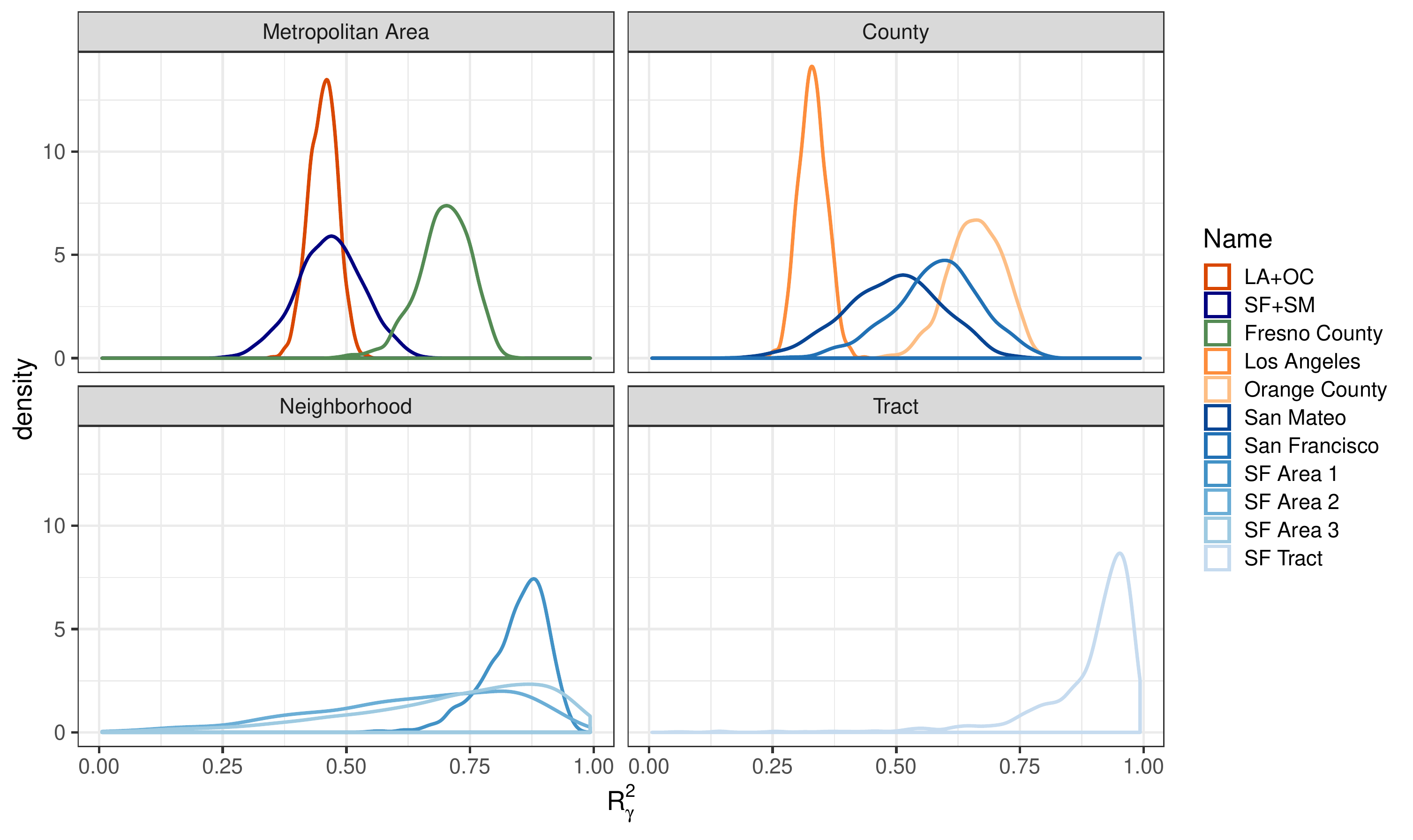}
  \caption{\label{fig:diagnostics-local} Diagnostics for local linear
    fits, at different levels of geographic aggregation: metropolitan
    areas, county, neighborhood, and tract.  Generally, as the summary
    type becomes more localized, $R^2_\gamma$ increases.  The
    exception is moving from Greater Los Angeles to Los Angeles
    County, which is so sprawling and heterogeneous that predictions
    of housing prices within cannot be distilled simply into a linear
    summary.  }
\end{figure}

Consider the fitted local linear summaries at the metropolitan area
level, shown in Figure~\ref{fig:selected-counties}.  As expected, the
relative importance of covariates does differ rather significantly
between the three areas.  For instance, population seems to positively
impact housing prices in the Bay Area, whereas household income has a
lower impact on housing prices there as compared to the two other
areas.  Interestingly, the summary predictive explainability for these
three areas differ widely, as shown in the top panel of
Figure~\ref{fig:diagnostics-local} which displays the $R^2_\gamma$
summary diagnostics.  Fresno has the high proportion of predictive
variation explained by the linear summary, while the LA area has the
lowest.  As we do not have observations at these generated predictive
locations $\tilde X$ for these locations, we do not report
$\phi_\gamma$ here, though this could also be calculated using draws
from the posterior predictive distribution
$p(\tilde y_i \mid Y, X, \tilde x_i)$.

We expect a greater degree of localization to lead to gains in
predictive explainability in the local linear summary.  While this is
true when comparing the $R^2_\gamma$ of the county-level linear
summary (coefficient estimates from which are not shown) for Orange,
San Francisco, and San Mateo Counties compared to those of their
respective encompassing metropolitan areas, the linear summary for Los
Angeles actually has lower predictive explainability than the
metropolitan-level summary.  This could potentially be due to the
sprawling nature of Los Angeles County---that trends in housing prices
there may simply be too complex to distill into a single linear
summary.

\begin{figure}[!htb]
  \centering
  \includegraphics[width=0.44\textwidth]{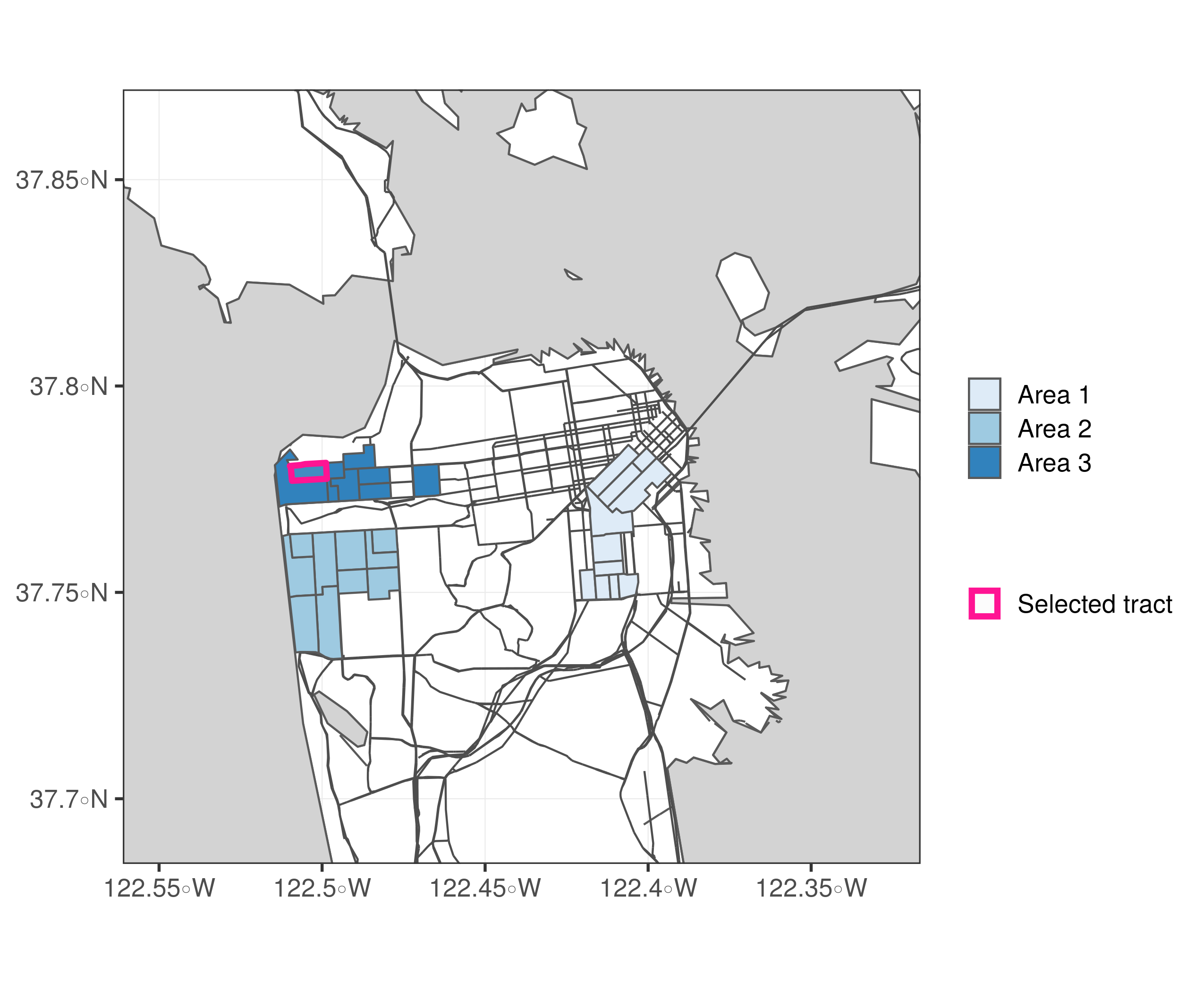}
  \includegraphics[width=0.55\textwidth]{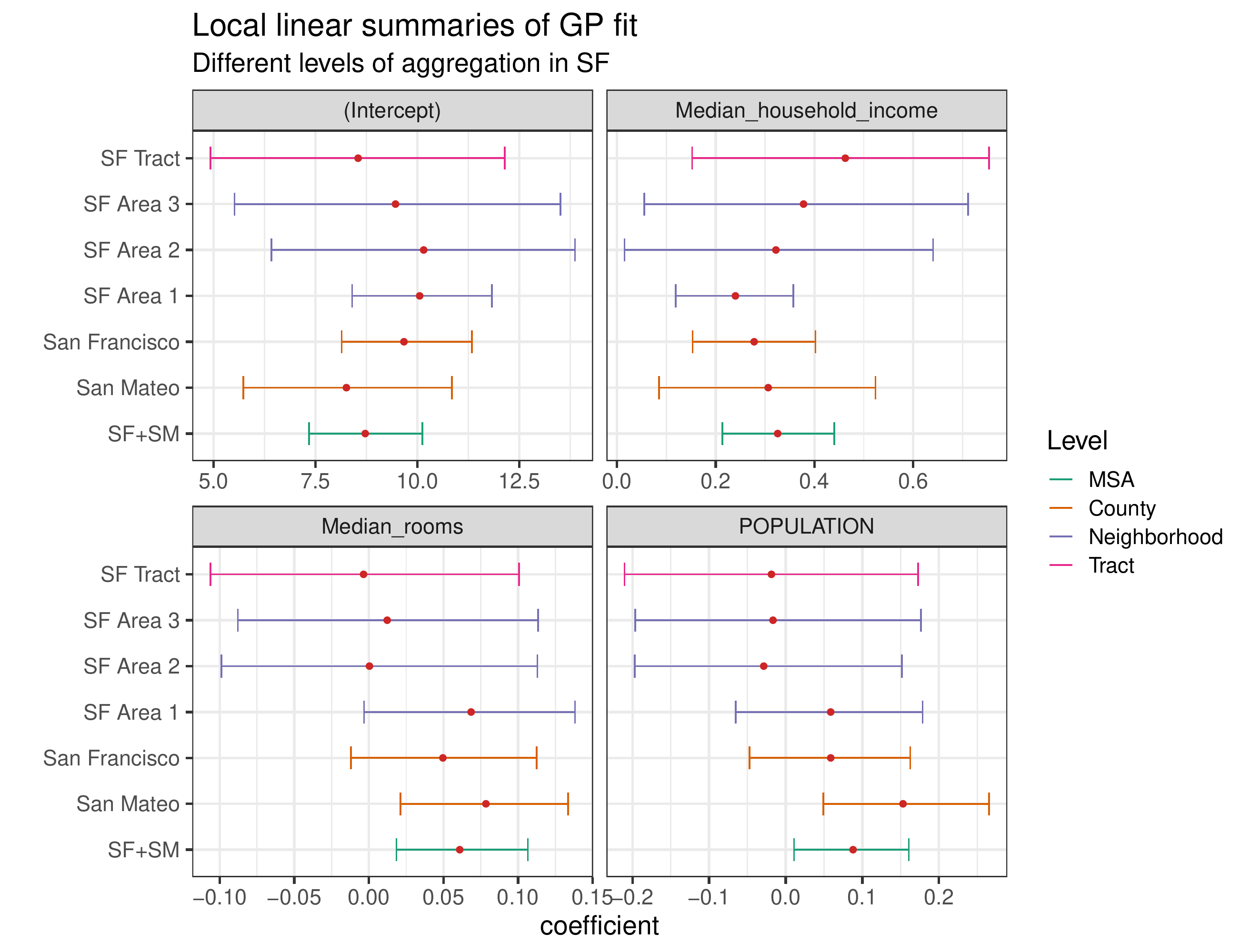}
  \caption{\label{fig:sf-areas} \textit{Left}: Selected areas for local
    linear fits in San Francisco, for three neighborhoods and one
    tract within one neighborhood.  \textit{Right}: Heterogeneity in
    local linear summaries within San Francisco, at different levels
    of geographic aggregation.  Note that at smaller areas of
    geographic aggregation, there is higher projected posterior
    variance. }
\end{figure}

We also consider three separate San Francisco neighborhoods, each
defined by sets of eight to twelve neighboring tracts, for which to
create local linear summaries.  We also create a model prediction
summary around a single selected tract located within one of these
neighborhoods.  Results for the these linear summaries, compared to
those from the encompassing metropolitan area and counties, are shown
in Figure~\ref{fig:sf-areas}.  Even within a relatively small-area
city like San Francisco there is marked variation in housing price
determinants.  Fittingly, there is greater projected posterior
variance in the smaller defined areas.  The combined panels of
Figure~\ref{fig:diagnostics-local} confirm our initial conjecture that
the predictive variation explained by summarization generally
increases for progressively local linear summaries.

\section{Discussion}
\label{sec:discussion}

When nonparametric models are used in regression analysis, there is a
natural tradeoff between model flexibility (and accuracy) and model
interpretability.  We attempt to bridge this gap, by separating model
specification and interpretation, using a two-stage approach that
yields valid Bayesian inference over multiple interpretable
quantities.   
This generalizes and expands upon previous work on posterior
summarization initiated by \cite{DSS} by introducing measures of
uncertainty via projected posteriors.  We also introduce tools for
explaining local variable importance, give metrics for checking the
quality of summaries, and provide heuristics for refining them as
necessary.  The approach outlined here is modular by design, allowing
for a wide array of summaries to be built for any suitably flexible
regression model, with any error distribution for the response.

The validity of these summaries is contingent upon having a good model
fit in the first stage.  If we do not regularize appropriately, then
we will fit to the noise in the first stage, and there will be
insufficient posterior variability in the summary.  If the fitted
model is otherwise biased, then the summary will similarly
misrepresent the true (unknown) regression function.  Therefore
standard model checks should be performed after the initial model is
fit.  As with any analysis, we are subject to fall victim to Simpson's
paradox if we do not carefully consider joint versus marginal trends.

In statistical inference, there is a distinction between confirmatory
analysis, where scientific hypotheses are specified \textit{a priori}
and then tested in light of the data, and exploratory analysis, where
data are used to generate hypotheses for future investigation.  Our
method falls somewhere between these two extremes.  Summaries will
typically be updated through the iterative process we describe, so
generally these analyses will not be confirmatory in the usual
sense. However, with our approach we do reduce researcher degrees of
freedom.  Instead of fitting and refining multiple models, and using
the data each time the best fitting one, we use the data only once to
find the best flexible estimate of the regression function without
regard to inference.  Thereafter, the fitted posterior is investigated
until an appropriate interpretable summary is found, thus resolving
the problem of ``posterior hacking.''


A closely related line of research is projective model selection for
generalized linear models \citep{GoutisRobert, DupuisRobert,
  Piironen2017, piironen2016projection
 }.  Under this
approach, the posterior for a full ``reference model'' is calculated,
and projected nested models are found by minimizing the
Kullback-Leibler divergence between predictive distributions of these
two models.  
The emphasis in these works is model selection, whereas our focus is
on giving interpretable explanations of models using a decision
theoretic approach.  However, this 
can be considered a special case of our procedure when this KL
divergence is used as the predictive discrepancy function in the
summary loss function.  \response{Further, \cite{Piironen2017,
    piironen2016projection} use this approach to rank variables, but
  to our knowledge this does not communicate the degree of
  nonlinearity or interaction effects leading to this ranking, which
  our method attempts to answer. }

\response{The calculation of the linear summary projection approach is
  quite similar to the ``effect size analog'' developed by
  \cite{crawford2018kernel,crawford2019predictor}, who also aim to
  quantify the influence of individual explanatory variables in
  nonlinear kernel models by projecting the nonlinear function onto
  the original covariate space.  They even propose to obtain a
  projected posterior using Monte Carlo posterior draws of the
  regression function, and then focus on variable selection after
  obtaining this projected posterior.  Our method explicitly specifies
  summarization as a decision problem, and embeds linear summaries
  into a broader class of available model summaries.  Furthermore, we
  propose to find a sparse linear summary by enforcing sparsity in the
  point estimate for the linear model summary via a penalty term, and
  then project posterior uncertainty onto this subset of coefficients,
  as opposed to selecting variables after finding the complete
  posterior for the linear projection with all variables. }

Additionally, our work is related to the field of interpretable
machine learning, where there has been much recent development.
Partial dependence plots \citep{Friedman2001}, and related tools like
individual conditional expectation plots \citep{ICE} and accumulated
local effects plots \citep{ALE} attempt to explain the partial effects
of individual covariates for generic black box models.  \response{To
  estimate the partial effect of a covariate $x_j$, these methods
  calculate the value of $f(x_j', x_{-j})$ over varying levels of
  $x_j'$, each time marginalizing over all other covariates $x_{-j}$.
  These methods have several drawbacks.  The resolution of the grid of
  $x_j'$ values must be specified, querying the model for so many
  iterations often requires significant computation time, and it is
  unclear how to propagate model uncertainty.  Instead, our method
  more directly seeks to characterize the predictive trends in $f$
  within a given region of covariate space by specifying the class of
  summary functions.  If we enforce this class of summaries to be
  additive, for example, then it allows us to make statements of
  average partial effects. In addition, fitting these
  lower-dimensional surrogates usually requires much less computation
  time compared to partial dependence plots, and if we have posterior
  draws for the vector of predictions $\mathbf{f}$, then it is
  efficient to calculate the projected posterior. }

Similar to our explanations of local model behavior, \cite{lime}
introduce the LIME method, which builds a local surrogate model to
explain individual predictions by the presence or absence of certain
binary features.  This method also repeatedly queries the output of
the fitted model.  In contrast, we calculate summaries by fitting
surrogate functions to the output of the model only at specified
locations.  Additionally, our calculated partial effects for both
local and global summaries are accompanied by valid uncertainty
estimates, and we quantify how well the summaries represent the
original model.

\response{The are several possible downsides to our approach.  Our
  method requires the use of a Bayesian model in order to characterize
  a model summary as a well-defined decision problem, and further to
  have rich understanding of how well the summary approximates the
  estimated regression function.  Of course, this almost always
  requires posterior sampling which can be computationally difficult
  with large sample sizes.  In addition, our interaction search
  procedure may break down in the presence of many covariates, say
  $p>100$, in which case the number of possible interactions becomes
  large.  One possible approach would be to use a first stage
  covariate selection, followed by a secondary exploratory stage,
  though it becomes more difficult to pose these two stages jointly as
  a single decision problem.  This avenue is left to future work.  }

Because of the generality of our developed approach, there is much
room for expanding this work.  Here we considered only a limited
number of potentially many possible model summaries.  We find the
prospect of applying this approach to other nonparametric models used
in different applications be very promising.  In particular, we plan
to produce interpretable summaries of nonparametric models for
heterogeneous treatment effect estimation.














\appendix





\section{Additional plots}
\label{sec:additional-plots}
\begin{figure}[htp!]
  \centering
  \includegraphics[width=0.99\textwidth]{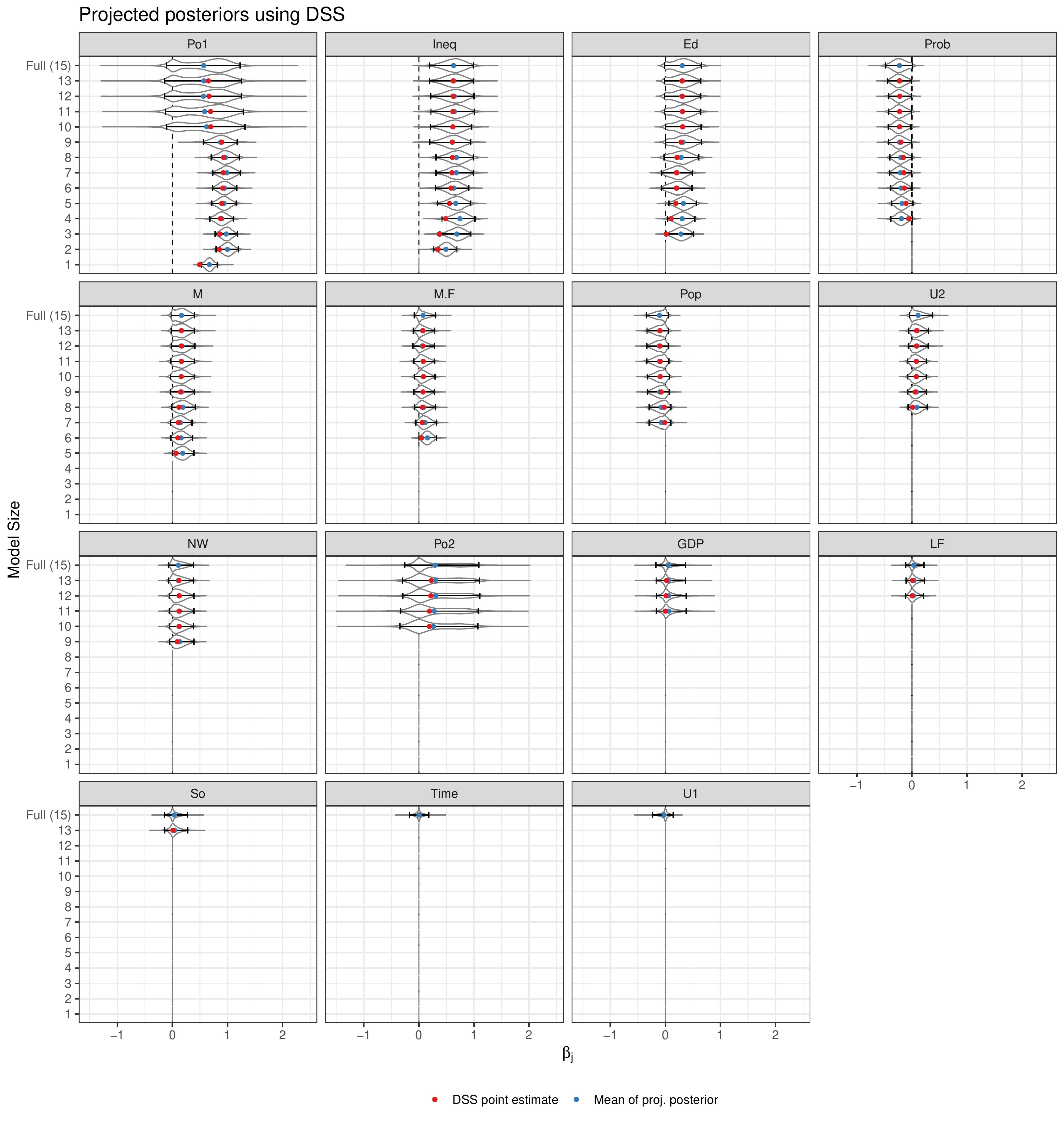}
  \caption{\label{fig:dss-crime-all-posteriors} All projected
    posteriors for US crime example in Section 2.1.  }
\end{figure}

\begin{figure}[ht]
  \centering
  \includegraphics[width=0.8\textwidth]{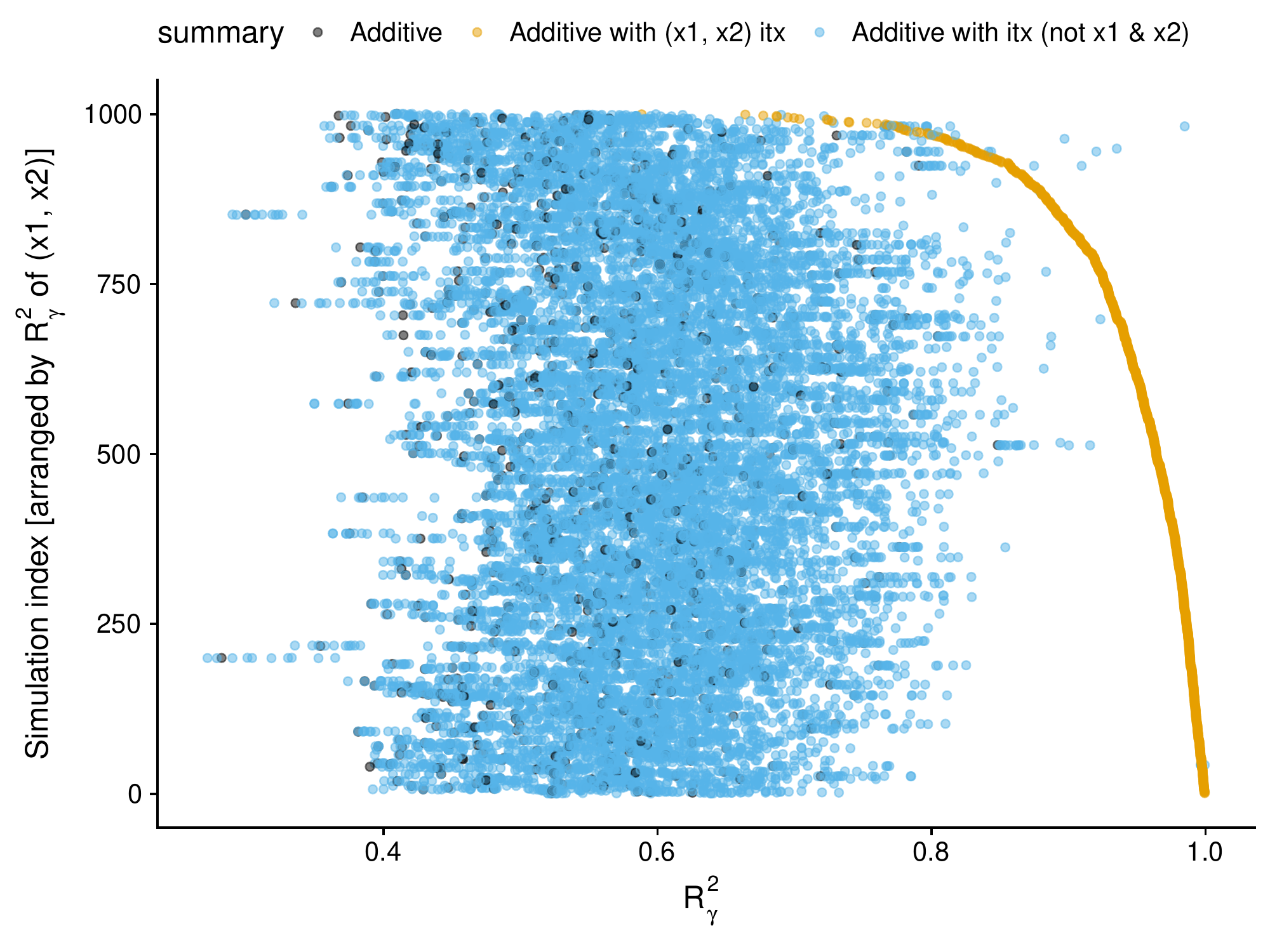}
  \caption{\label{fig:rsq-sims-big} A series of 1000 replications of
    the simulation example in Section 4.2.  Here we show gains in
    $R^2_\gamma$ over the purely additive summary from fitting
    partially additive summaries with a single one-way interaction.  }
\end{figure}

\begin{figure}[!htb]
  \centering
  \includegraphics[width=0.8\textwidth]{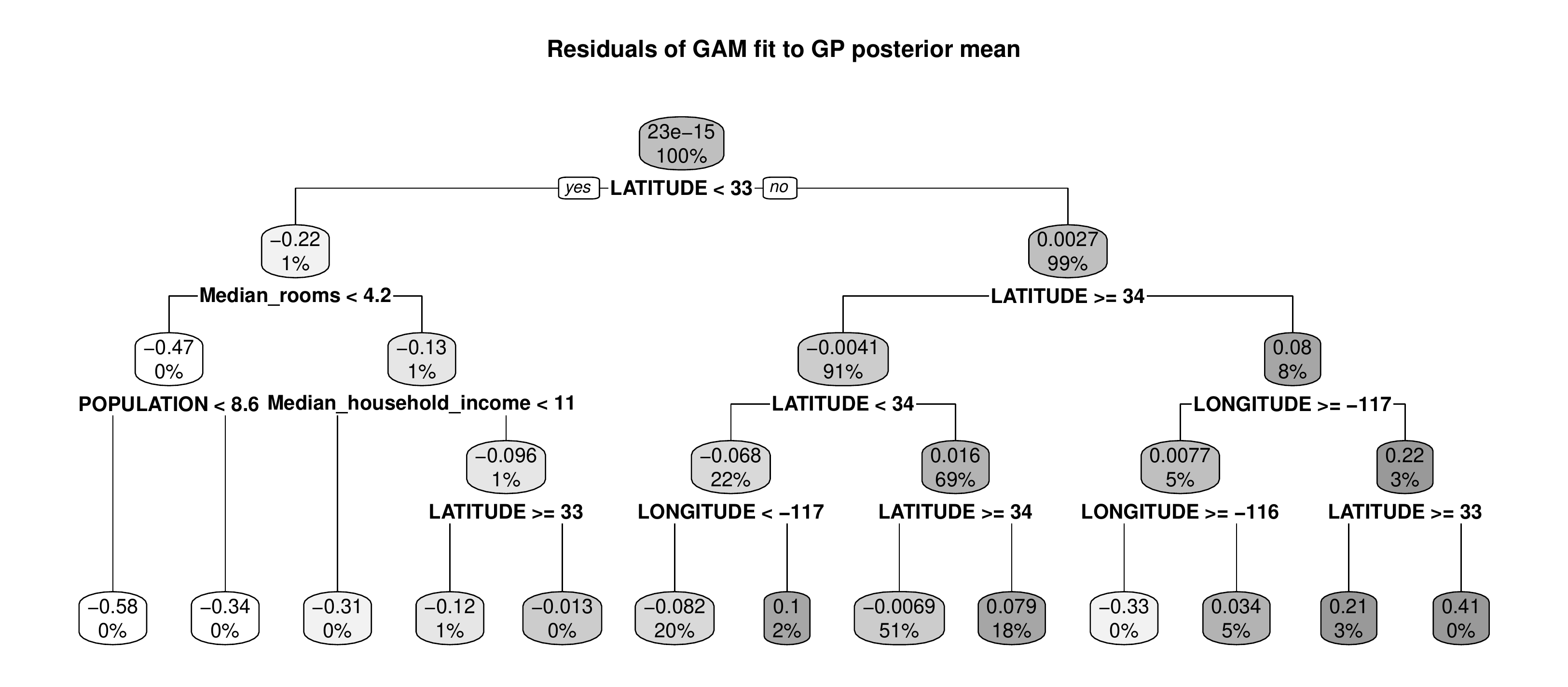}
  \includegraphics[width=0.8\textwidth]{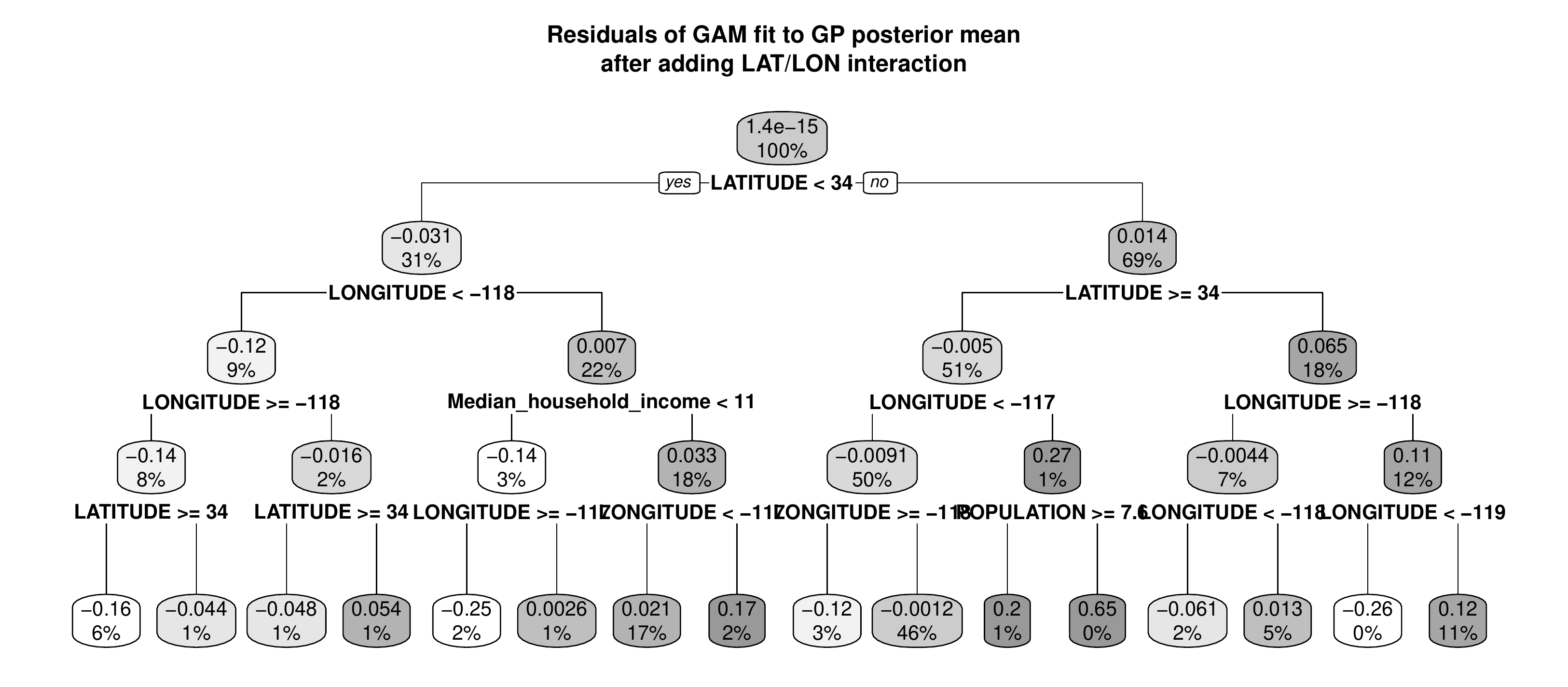} 
  \caption{\label{fig:residual-cart} Regression trees fit to
    summarization residuals from the purely additive model
    summarization of GP fit, before \textit{(top)} and after
    \textit{(bottom)} adding a spatial interaction term for the case
    study in Section 5.  These plots serve as diagnostics to give us a
    clue of important interactions not yet taken into account.  }
\end{figure}

See Figures~\ref{fig:dss-crime-all-posteriors},  \ref{fig:rsq-sims-big}, and
\ref{fig:residual-cart}. 

\bibliographystyle{plainnat}

\singlespacing
  \bibliography{main}  


\end{document}